\documentclass[aps,prb,reprint,showpacs]{revtex4-1}
\usepackage{amsfonts}
\usepackage{amsmath}
\usepackage{amssymb}
\usepackage[pdftex]{graphicx}
\begin{document}
\title{Stability of dynamic coherent states in intrinsic Josephson-junction stacks
near internal cavity resonance}
\author{A. E. Koshelev}
\affiliation{Materials Science Division, Argonne National Laboratory, Argonne, Illinois 60439}
\date{\today}

\begin{abstract}
Stacks of intrinsic Josephson junctions in the resistive state can
by efficiently synchronized by the internal cavity mode resonantly
excited by the Josephson oscillations. We study the stability of
dynamic coherent states near the resonance with respect to small
perturbations. Three states are considered: the homogeneous and
alternating-kink states in zero magnetic field and the homogeneous
state in the magnetic field near the value corresponding to half
flux quantum per junction. We found two possible instabilities
related to the short-scale and long-scale perturbations. The
homogeneous state in modulated junction is typically unstable with
respect to the \emph{short-scale alternating phase deformations}
unless the Josephson current is completely suppressed in one half of
the stack. The kink state is stable with respect to such
deformations and homogeneous state in the magnetic field is only
stable within a certain range of frequencies and fields.  Stability
with respect to \emph{the long-range deformations} is controlled by
resonance excitations of fast modes at finite wave vectors and
typically leads to unstable range of the wave-vectors. This range
shrinks with approaching the resonance and increasing the in-plane
dissipation. As a consequence, in finite-height stacks the stability
frequency range near the resonance increases with decreasing the
height.
\end{abstract}
\maketitle

\section{Introduction}

Superconducting tunneling junctions are natural voltage-tunable
sources of electromagnetic radiation due to the \textit{ac} Josephson
effect.\cite{JosPL62} As radiation from a single junction is very
small, large-size arrays of artificially fabricated junctions have
been used to enhance power of electromagnetic radiation, see early
reviews \cite{Jain,Darula99} and more recent
papers.\cite{BarbaraPRL99,SongAPL09} The main challenge is to
synchronize all junctions in the array. In this case the total
emitted power is expected to be proportional to the square of the total
number of junctions.

Intrinsic Josephson junctions in the high-temperature layered
superconducting materials\cite{KleinerPRL92}, such as
Bi$_{2}$Sr$_{2}$CaCu$_{2}$O$_{8} $ (BSCCO), provide a very promising
base for developing coherent generators of electromagnetic radiation
which may operate in the terahertz frequency range. These materials
have several important advantages in comparison with artificial
structures made out of conventional superconductors including (i)
the large packing density of the junctions, (ii) a large value of
the superconducting gap (up to 60 mev) which allows to bring the
Josephson frequency into the terahertz range, and (iii) possibility
to make very large arrays of practically identical junctions.

The stack of junctions can be a powerful, coherent, and efficient
generator only if the oscillations of the superconducting phases in
all junctions are synchronized. Due to weak intrinsic interaction
between the junctions, this is a challenging task. One possible way
to synchronization is to use interactions between the junctions via
the generated external radiation \cite{BulKoshPRL07}. In this case,
for efficient coupling to the radiation field, a junction stack
(mesa) must have small lateral size ($<10$ $\mu$m) and contain a
very large number of junctions ($>10^{4}$). Such a mesa would be a
frequency-tunable source with the considerable power conversion
efficiency. The obvious technological challenge of this design is
the requirement to fabricate structures with such large number of
almost identical junctions. This design has not yet been implemented
in practice.

A very promising route to efficient synchronization is to excite an
internal cavity resonance in finite-size samples
(mesas).\cite{LutfiSci07,KoshBulPRB08} Being excited, the resonance
mode can entrain oscillations in a very large number of junctions.
The frequency of this mode is set by the lateral size of the mesa
and for the resonance frequency in the terahertz range the width has
to be rather large (40-100 $\mu$m). The experimental demonstration
\cite{LutfiSci07} and confirmation \cite{KadowakiTHz,WangKleiner} of
this mechanism marks a major advance in the quest for Josephson
terahertz generators.

In general, the structure of the coherent state and the mechanism of pumping energy
into the cavity mode are nontrivial issues. Homogeneous phase oscillations at
zero magnetic field do not couple to the Fiske modes. Such coupling can be
facilitated by introducing an external modulation of the Josephson critical
current density.\cite{KoshBulPRB08} In this case the amplitudes of the
generated standing wave and of the produced radiation are proportional to the
strength of the modulation.

An interesting alternative possibility has been demonstrated
recently.\cite{LinHuPRL08,AltStatePRB,KoshBulLT25} It was found that
near the resonance an inhomogeneous synchronized state is formed. In
this state, the stack spontaneously splits into two subsystems with
different phase-oscillation patterns, corresponding to the
alternating phase kinks and antikinks \emph{statically} located near
the center. This leads to a static phase shift between the
oscillations in the two subsystems varying from 0 to 2$\pi$ in a
narrow region near the stack center. In spite of this c-axis
inhomogeneity, the oscillating electric and magnetic fields are
almost homogeneous in all the junctions. The formation of this state
promotes efficient pumping of the energy into the cavity resonance.

Another potential candidate for the coherent state producing strong
emission is a homogeneous state in the external magnetic field,
also known as a rectangular Josephson-vortex lattice. In spite of strong
experimental efforts, the existence of this state has not yet been clearly
demonstrated, except for small-size stacks at very small
velocity.\cite{RectSmallStacks} It was argued that in the large-size
crystals the rectangular lattice is almost always
unstable.\cite{ArtemRemPRB03} For a finite-size system the stability
analysis of the homogeneous state has been performed recently
\cite{RakhPRB09} and stability regions have been found.

As large-size stacks have a huge number of degrees of freedom,
stability of the coherent states with respect to small perturbations
is an important and nontrivial issue. Linear stability analysis
amounts to calculating the full frequency spectrum for small
perturbations with respect to steady-state coherent solutions and
verifying that there are no exponentially growing perturbations. The
stability analysis allows to evaluate the range of parameters where
the coherent states are possible. For linear arrays of point
junctions the stability analysis has been performed in Refs.
\onlinecite{HadleyPRB88,ChernikovPRB95}. In this case stability is
strongly influenced by the external load. A large array of the
small-size intrinsic Josephson junctions can be stabilized by the
radiation field which acts similar to the external capacitive
load.\cite{BulKoshPRL07}

In this paper we perform a systematic comparative analysis of the
linear stability of different coherent states in the array of
extended Josephson junctions near the resonance. We revealed two
types of instabilities. The \emph{short wave-length} instability corresponding to the alternating phase deformations
develops for states which have regions of negative local
time-averaged Josephson coupling. This instability is sensitive to
the nature of the dynamic state. The homogeneous state in the
modulated junctions is typically prone to this kind of instability.
The \emph{long wave-length} instability with wave lengths larger than the London penetration depth $\lambda$ appears due to the
parametric resonance excitation of the fast modes at finite wave
vectors. Analysis of this instability is essentially identical for
all dynamic states. The instability criterion depends on several
factors including behavior of the resonance frequency shift with
increasing the in-plane and c-axis wave vectors and the relation between
damping of the uniform mode and modes at finite wave vectors. The
most essential parameters influencing stability include the shift of
the Josephson frequency with respect to the resonance, the stack height,
and the in-plane quasiparticle dissipation. The paper is organized as
follows. In Section \ref{Sec:EqSol} we present the dynamic phase
equations and coherent solutions near the internal resonance. In
Section \ref{Sec:EqPert} we derive the linear dynamic equations for
small perturbations with respect to steady states. In Section
\ref{Sec:ShortWave} we consider the short-wave length instabilities
of different steady states and present the numerical test for these
instabilities in the stacks with modulated critical current. In
Section \ref{Sec:LongWave} we describe the stability analysis with
respect to the long-wave deformations and numerical verification of
this analysis. The description of numerical simulations used to
check some of our analytical results is presented in Appendix
\ref{App-Num}.

\section{Phase dynamic equations and coherent solutions\label{Sec:EqSol}}

The dynamic equations for the Josephson-junction stacks have been
derived in several papers \cite{DynamEqs} and have been used in
different forms in numerous simulation and theoretical
studies.\cite{SimulJJ} These equations can be written in the reduced
form as coupled equations for the phases $\varphi_{n}$ and
dimensionless magnetic fields $\mathbf{h}_{n}=(h_{x,n} ,h_{y,n})$,
\begin{align}
&\frac{\partial^{2}\varphi_{n}}{\partial\tau^{2}}\!+\!\left(  1\!-\!\alpha
\nabla_{n}^{2}\right)  \left(  \nu_{c}\frac{\partial\varphi_{n}}{\partial\tau
}\!+\!g(x)\sin\varphi_{n}\!-e_{ijz}\partial_{i}h_{j,n}\right)  \!
=\!0,\label{EqPhase}\\
&\ell^{2}\nabla_{n}^{2}h_{j,n}\!-\!\left(  1\!+\!\nu_{ab}\frac{\partial
}{\partial\tau}\right)  \left[  h_{j,n}-e_{ijz}\partial_{i}\varphi_{n}\right]
  =\!0. \label{EqField}
\end{align}
Here $i,j=x,y$, $e_{ijz}$ is the Levi-Civita symbol
($e_{xyz}=-e_{yxz} =1$). The units in these equations are selected
as follows: $1/\omega_{p}$ is the unit of time with $\omega_{p}$
being the plasma frequency, the c-axis London penetration depth
$\lambda_{c}$ is the unit of length, $\Phi_{0}/(2\pi \lambda_{c}s)$
is the unit of the magnetic field. The function $g(x)$ describes
possible modulation of the Josephson current which was suggested as
the way to couple to the internal resonance.\cite{KoshBulPRB08} We
consider both modulated and unmodulated ($g(x)\!=\!1$) cases. The
equations depend on four parameters, the layer-charging parameter
$\alpha$,\cite{Koyama}   two damping parameters,
$\nu_{c}=4\pi\sigma_{c}/(\varepsilon_{c}\omega_{p} )$,\
$\nu_{ab}=4\pi\sigma_{ab}/(\varepsilon_{c}\gamma^{2}\omega_{p})$,\
and the ratio $\ell=\lambda/s$, where $\sigma_{c}$ and $\sigma_{ab}$
are the components of quasiparticle conductivity, $\lambda$ is the
in-plane London penetration depth, and $\gamma$ is the anisotropy
factor. We will study a finite-size stack (mesa) containing $N$
junctions with lateral sizes $L_{x}$ and $L_{y}$. We consider the
case $L_{y}\gg L_{x}$.

We tested some of our analytical results with numerical simulations.
The numerical procedure is the same as in Ref.\
\onlinecite{AltStatePRB}. For completeness, details of the numerical
simulations are described in Appendix \ref{App-Num}.

\subsection{Coherent states in zero magnetic field}

We consider the stack in the coherent resistive state, in which all junctions
have identical voltage drops. In this state the dynamical phase can be written
as\cite{footnoteSelfField}
\[
\varphi_{n}(x,\tau)\approx\omega\tau+\theta_{n}(x,\tau).
\]
Further, we assume that the Josephson frequency $\omega$ is close to
the in-phase resonance frequency $\omega_{m}=m\pi/L_{x}$ and the
resonance cavity mode is excited inside the mesa meaning that the
oscillating phase has the large resonance contribution
$\theta_{n}(x,\tau)\sim\cos(m\pi x/L_{x} )\cos(\omega\tau+\alpha)$.
We will mostly focus on the experimentally relevant case of the
fundamental mode $m=1$. One can distinguish two particular cases:
for a homogeneous (uniform) solution the phases $\theta_{n}(x,\tau)$
are identical in all junctions and for an inhomogeneous (nonuniform)
solution the phases $\theta_{n} (x,\tau)$ vary from junction to
junction. For the homogeneous solution \cite{KoshBulPRB08},
$\theta_{n}(x,\tau)=\theta(x,\tau)$, Eqs.\ (\ref{EqPhase} ) and
(\ref{EqField}) reduce to the sine-Gordon equation,
\begin{equation}
\frac{\partial^{2}\varphi}{\partial\tau^{2}}+\nu_{c}\frac{\partial\varphi
}{\partial\tau}+g(x)\sin\varphi-\frac{\partial^{2}\varphi(x,\tau)}{\partial
x^{2}}=0.
\end{equation}
In this case coupling to the resonance mode is only induced by the external
modulation $g(x)$. Representing the oscillating phase as $\theta(x,\tau
)=\operatorname{Re}[\theta_{\omega}(x)\exp(-i\omega\tau)]$, we obtain an equation
for the complex amplitude $\theta_{\omega}(x)$,
\begin{equation}
\left(  \omega^{2}+i\nu_{c}\omega\right)  \theta_{\omega}+\frac{\partial
^{2}\theta_{\omega}}{\partial x^{2}}=ig(x). \label{ComplAmpl}
\end{equation}
This equation has to be supplemented by the boundary conditions accounting for
radiation. For symmetric mesas general boundary conditions can be presented in
the following form
\begin{subequations}
\label{BoundCond}
\begin{align}
\frac{\partial\theta_{\omega}}{\partial x}(L_{x})  &  =i\zeta\theta_{\omega
}(L_{x})+i\tilde{\zeta}\theta_{\omega}(0),\\
\frac{\partial\theta_{\omega}}{\partial x}(0)  &  =-i\zeta\theta_{\omega
}(0)-i\tilde{\zeta}\theta_{\omega}(L_{x})
\end{align}
The coefficients $\zeta$ and $\tilde{\zeta}$ depend on the particular geometry. For example, for the case of an isolated mesa on a metallic plate with thin
metallic contact on the top
\end{subequations}
\begin{align*}
\zeta &  \approx\frac{\omega^{2}L_{z}}{2\varepsilon_{c}}\left(  1-\frac
{2i}{\pi}\ln\left(  \frac{C}{k_{\omega}L_{z}}\right)  \right)  ,\\
\tilde{\zeta}  &  \approx-\frac{\omega^{2}L_{z}}{2\varepsilon_{c}}\left[
J_{0}(k_{\omega}L_{x})+iN_{0}(k_{\omega}L_{x})\right]  ,
\end{align*}
where $k_{\omega}=\omega/c$, $L_{z}=Ns$ is the stack height, $J_{0}(z)$ and
$N_{0}(z)$ are the Bessel functions, and $C\sim1$. As $\zeta$ and
$\tilde{\zeta}$ are small, near the resonance we can look for solution in the
form
\[
\theta_{\omega}=\psi\cos(m\pi x/L_{x})+\upsilon(x)
\]
where $\upsilon(x)$ is the small correction accounting for the
radiation boundary conditions. With this ansatz, Eq.
(\ref{ComplAmpl}) becomes
\begin{align*}
&  \left(  \omega^{2}\!-\!\omega_{m}^{2}\!+\!i\nu_{c}\omega\right)  \psi
\cos\left(  m\pi x/L_{x}\right)  \!\\
&  +\!\left(  \omega^{2}\!+i\nu_{c}\omega\right)  \upsilon+\frac{\partial
^{2}\upsilon}{\partial x^{2}}\!=\!ig(x)
\end{align*}
As $\upsilon$ is small, we can approximately replace $\left(
\omega^{2} +i\nu_{c}\omega\right)
\upsilon\rightarrow\omega_{m}^{2}\upsilon$ and look for correction
satisfying the mode-matching condition $\omega_{m}^{2}
\upsilon+\partial^{2}\upsilon/\partial x^{2}\propto\cos(m\pi
x/L_{x})$. In this case, $\upsilon(x)$ can be found as
\[
\upsilon(x)\approx a\psi\ (x-L_{x}/2)\sin(m\pi x/L_{x})
\]
giving $\omega_{m}^{2}\upsilon+\partial^{2}\upsilon/\partial x^{2}=2a\psi
\cos(m\pi x/L_{x})\left(  m\pi/L_{x}\right)  .$ Substituting this result into
the previous equation and taking projection to the mode, we obtain
\[
\psi=\frac{ig_{m}}{\omega^{2}-\omega_{m}^{2}+2a\left(  m\pi/L_{x}\right)
+i\nu_{c}\omega}
\]
where $g_{m}=(2/L_{x})\int_{0}^{L_{x}}\cos(m\pi x/L_{x})g(x)dx$ is the
coupling parameter. The complex constant $a$ has to be found from the boundary
conditions (\ref{BoundCond}) where in the righthand side we can neglect
$\upsilon(x)$. In this case the approximate boundary conditions for the
correction become
\begin{align*}
\frac{\partial\upsilon}{\partial x}(L_{x})  &  \approx i\left[  (-1)^{m}
\zeta+\tilde{\zeta}\right]  \psi\\
\frac{\partial\upsilon}{\partial x}(0)  &  \approx-i\left[  \zeta
+(-1)^{m}\tilde{\zeta}\right]  \psi
\end{align*}
and they give identical results for $a$,
\[
a\approx\left(  2i/m\pi\right)  \left[  \zeta+(-1)^{m}\tilde{\zeta}\right]  .
\]
The amplitude of the cavity mode can finally be represented as
\begin{equation}
\psi=\frac{ig_{m}}{(1+\alpha_{r})\omega^{2}-\omega_{m}^{2}+i\left(  \nu
_{r}+\nu_{c}\right)  \omega} \label{ModeAmplit}
\end{equation}
where
\[
\nu_{r}=\frac{4}{\omega L_{x}}\operatorname{Re}\left[
\zeta+(-1)^{m} \tilde{\zeta}\right]  =\frac{2\omega
L_{z}}{\varepsilon_{c}L_{x}}\left[
1-(-1)^{m}J_{0}(k_{\omega}L_{x})\right]
\]
determines radiation contribution to the damping
\cite{KoshBulPRB08,footnote} and
\begin{align*}
\alpha_{r}  &  =-\frac{4}{L\omega^{2}}\operatorname{Im}\left[  \zeta
+(-1)^{m}\tilde{\zeta}\right] \\
&  \approx \frac{L_{z}}{\pi\varepsilon_{c}L_{x}}\left[  \ln\left(
\frac{C} {k_{\omega}L_{z}}\right)
+(-1)^{m}\frac{\pi}{2}N_{0}(k_{\omega}L_{x})\right]
\end{align*}
determines the resonance frequency shift due to radiation. This
small frequency shift is frequently neglected. However, it will be
essential for the long-range stability analysis. Therefore, the
homogeneous oscillating phase near the resonance can be represented
as
\begin{equation}
\theta(x,\tau)\!\approx\!-\operatorname{Im}\left[  \frac{g_{m}\exp
(-i\omega\tau)}{(1\!+\!\alpha_{r})\omega^{2}\!-\!\omega_{m}^{2}\!+\!i\nu
\omega}\right]  \cos\left(  \frac{m\pi x}{L_{x}}\right),
\label{HomogPhase}
\end{equation}
where $\nu=\nu_{c}+\nu_{r}$ is the total mode-damping
parameter.\cite{KoshBulPRB08}

The homogeneous solution is not the only possible coherent state. A
spectacular example of an inhomogeneous coherent solution is the
alternating-kink state recently reported in
Refs.\ \onlinecite{LinHuPRL08,AltStatePRB}. For this state the phase
distribution is given by,
\[
\theta_{n}(x,\tau)=(-1)^{n}\theta_{k}(x)+\theta(x,\tau),
\]
where, for the fundamental mode, the static-kink phase
$\theta_{k}(x)$ changes from $0$ to $\pi$ near the center. As the
region of this change is extremely narrow \cite{AltStatePRB}, in
the equation for the homogeneous oscillating phase $\theta(x,\tau)$ one
can approximate $\theta_{k}(x)$ with the step function
$\theta_{k}(x)\rightarrow\pi\ \Theta(x-L_{x}/2)$, where
$\Theta(x)=0$ for $x<0$ and $\Theta(x)=1$ for $x>0$. Within this
approximation, such a state becomes equivalent to the stack with
modulated current density with the modulation function
$g_{k}(x)=\mathrm{sgn}(x-L_{x}/2)$. In this case the homogeneous
part of the oscillating phase is again given by Eq.\
(\ref{HomogPhase}), where for $m=1$, $g_{m}\rightarrow g_{k,1}
\approx4/\pi$. While for the homogeneous solution coupling to the
resonance mode only exists due to the external modulation, for the
alternating-kink state such coupling is self-generated.

\subsection{Homogeneous solution in magnetic field}

We also consider the homogeneous resonance solution induced by the external
magnetic field $h_{e}$ applied along $y$-direction,
\[
\varphi(x,\tau)=\omega\tau+h_{e}x+\theta(x,\tau)
\]
in a stack without the external modulation, $g(x)=1$. The
oscillating phase
$\theta(x,\tau)=\operatorname{Re}[\theta_{\omega}(x)\exp(-i\omega\tau)]$
is determined by the following equation
\[
\left(  \omega^{2}+i\nu_{c}\omega\right)  \theta_{\omega}+\frac{\partial
^{2}\theta_{\omega}}{\partial x^{2}}=i\exp(-ih_{e}x)\text{.}
\]
We will mainly focus on the most interesting case of the fundamental
mode and magnetic fields corresponding to a magnetic flux through
each junction close to half flux quantum, $h_{e}=\pi/L_{x}$,
providing the most efficient coupling to this resonance. The
homogeneous solution for the junction stack is essentially identical
to the corresponding solution for a single junction \cite{Kulik65},
see also Refs.\ \onlinecite{BulKoshPRL06,RakhPRB09}. In particular,
when the frequency is close to the resonance frequency $\omega_{m}$,
the dominating contribution to the oscillating phase again has the
resonance form given by Eq.\ (\ref{HomogPhase}) with
\begin{align}
g_{m}\!  &  \rightarrow g_{h,m}\!=\!\frac{2}{L_{x}}\int_{0}^{L_{x}}
\!dx\cos(m\pi x/L_{x})\exp(-ih_{e}x)\nonumber\\
&  =-\frac{2i\left(  1-(-1)^{m}\exp\left[  -ih_{e}L_{x}\right]  \right)
h_{e}L_{x}}{\left(  h_{e}L\right)  ^{2}-\left(  m\pi\right)  ^{2}}
\label{CouplParH}
\end{align}
being the coupling parameter due to the magnetic field. In contrast to the
zero-field case, this parameter is a complex number. For the fundamental mode
\begin{equation}
g_{h,1}=\frac{4ih_{e}L_{x}\exp\left[  -ih_{e}L_{x}/2\right]  \cos\left[
h_{e}L_{x}/2\right]  }{\pi^{2}-h_{e}^{2}L_{x}^{2}}.
\label{FieldCoupl}
\end{equation}
In particular, $g_{h,1}=1$ for $h_{e}L_{x}=\pi$. This function is
shown in the upper left plot of Fig.\ \ref{Fig-FieldPlots}.

\section{Equations for small perturbations with respect to coherent dynamic
state\label{Sec:EqPert}}

We will study the linear stability of the homogeneous solution with
respect to small perturbations. A similar analysis has been done in
a recent paper \cite{RakhPRB09} for the homogeneous solution in
magnetic field. We found, however, several important instabilities
which were missed in this paper.

We consider first the case of the homogeneous state in modulated
junctions.  With very good accuracy this analysis can also be applied to the kink state, because c-axis inhomogeneities in this state are located in a very
narrow region near the center. Perturbing the homogeneous
solution, $\varphi_{n}(x,\tau)=\omega\tau
+\theta(x,\tau)+\vartheta_{n}(\mathbf{r},\tau)$ and $\mathbf{h}_{n}
(\mathbf{r},\tau)=\mathbf{h}_{n}^{(0)}(x,\tau)+\tilde{\mathbf{h}}
_{n}(\mathbf{r},\tau)$, we obtain the linear equations for small
perturbations $\vartheta_{n}(\mathbf{r},\tau)$ and
$\tilde{\mathbf{h}}_{n}(\mathbf{r},\tau )$,
\begin{align}
&  \frac{\partial^{2}\vartheta_{n}}{\partial\tau^{2}}+\left(  1\!-\!\alpha
\nabla_{n}^{2}\right) \nonumber\\
&  \times\left(  \nu_{c}\frac{\partial\vartheta_{n}}{\partial\tau
}\!+\!g(x)C(x,\tau)\vartheta_{n}\!-\!e_{ijz}\partial_{i}\tilde{h}
_{j,n}\right)  =0,\label{EqSmallDevPh}\\
&  \ell^{2}\nabla_{n}^{2}\tilde{h}_{j,n}\!-\!\left(  1\!+\!\nu_{ab}
\frac{\partial}{\partial\tau}\right)  \left[  \tilde{h}_{j,n}-e_{ijz}
\partial_{i}\vartheta_{n}\right]  =\!0, \label{EqSmallDevH}
\end{align}
where $C(x,\tau)\equiv\cos\left[  \omega\tau+\theta(x,\tau)\right]  $. Due to
this oscillating cosine, perturbations are not monochromatic, oscillations
with the frequency $\Omega$ are coupled with the frequencies $\Omega\pm\omega
$. At $\omega\gg1$ we can look for solution in the form
\begin{align}
\vartheta_{n}(\mathbf{r},\tau)\!  &  \approx\!\sum_{q,k_{y}}\cos
[q(n+1/2)]\cos\left(  k_{y}y\right)  \exp\left(  -i\Omega\tau\right)
\nonumber\\
&  \times\sum_{\beta=0,\pm1}\!\tilde{\vartheta}_{\mathbf{k},\beta}
(x)\exp(i\beta\omega\tau),\label{PertPhase}\\
&  \text{with }q=\pi m_{z}/(N+1),k_{y}=\pi m_{y}/L_{y}\nonumber
\end{align}
and neglect other frequency components. For simple treatment of the
c-axis modes, we assumed that the stack is bounded by metallic
contacts which can be approximated by ideal
conductors.\cite{footnote} In this presentation $\Omega\equiv
\Omega(k_{y},q)$ is the complex eigenfrequency which has to be found
from equations (\ref{EqSmallDevPh}) and (\ref{EqSmallDevH}) and the
boundary conditions. The state is stable only if
$\operatorname{Im}[\Omega(k_{y},q)]<0$ for all $k_{y}$ and for all
$q$'s from $0$ to $\pi$. Separating the slow and fast parts in
equations (\ref{EqSmallDevPh}) and (\ref{EqSmallDevH}), using
relation
\begin{align*}
&  g(x)C(x,\tau)\vartheta_{n}(\mathbf{r},\tau)\approx\frac{g(x)}{2}
\!\sum_{q,k_{y}}\cos[q(n+1/2)]\cos\left(  k_{y}y\right) \\
&  \times\left[  \tilde{\vartheta}_{\mathbf{k},0}\sum_{\beta=\pm1}\exp
(i\beta\omega\tau-i\Omega\tau)\!+\!\sum_{\beta=\pm1}\!\tilde{\vartheta
}_{\mathbf{k},\beta}\exp(-i\Omega\tau)\right]  ,
\end{align*}
and, excluding the magnetic fields,
\[
\tilde{h}_{i,\mathbf{k},\beta}=\frac{1}{1+\ell^{2}\tilde{q}^{2}/\left[
1\!-\!i\nu_{ab}\left(  \Omega\!-\!\beta\omega\right)  \right]  }
e_{ijz}\partial_{j}\tilde{\vartheta}_{\mathbf{k},\beta},
\]
with $\beta=0,\pm1$ and $\tilde{q}=2\sin(q/2)$, we obtain coupled equations
for the slow and fast phase oscillations, see also Refs.
\onlinecite{ArtemRemPRB03,BulKoshPRL07,RakhPRB09},
\begin{align}
\left(  \tilde{\Omega}_{0}^{2}-g(x)\bar{C}-G_{q,0}^{-2}k_{y}^{2}\right)   &
\tilde{\vartheta}_{\mathbf{k},0}+G_{q,0}^{-2}\frac{\partial^{2}\tilde
{\vartheta}_{\mathbf{k},0}}{\partial x^{2}}\nonumber\\
&  =\frac{g(x)}{2}\sum_{\beta=\pm1}\tilde{\vartheta}_{\mathbf{k},\beta
},\label{EqSlow}\\
\left(  \tilde{\Omega}_{\beta}^{2}-G_{q,\beta}^{-2}k_{y}^{2}\right)
\tilde{\vartheta}_{\mathbf{k},\beta}+  &  G_{q,\beta}^{-2}\frac{\partial
^{2}\tilde{\vartheta}_{\mathbf{k},\beta}}{\partial x^{2}} =\frac
{g(x)\tilde{\vartheta}_{\mathbf{k},0}}{2}, \label{EqFast}
\end{align}
where we introduced notations
\begin{align*}
\tilde{\Omega}_{\beta}^{2}  &  \equiv\left(  \Omega-\beta\omega\right)
^{2}/(1+\alpha\tilde{q}^{2})+i\nu_{c}\left(  \Omega-\beta\omega\right)  ,\\
G_{q,\beta}^{2}  &  \!\equiv\!1+\ell^{2}\tilde{q}^{2}/[1\!-\!i(\Omega
-\beta\omega)\nu_{ab}].
\end{align*}
Using $\varphi=\omega\tau+\operatorname{Re}\left[  \theta_{\omega}
(x)\exp(-i\omega\tau)\right]  $, the time averaged cosine $\bar{C}
(x)\equiv\!\langle\cos\varphi\rangle_{\tau}$ can be evaluated as
\begin{align}
\bar{C}(x)\!  &\approx%
-\operatorname{Im}\left[  \theta_{\omega}(x)\right]/2\nonumber\\
&\approx-\frac{g_{m}\left[(1+\alpha_{r})\omega^{2}-\omega_{m}^{2}\right]}
{\left[(1\!+\!\alpha_{r})\omega^{2}\!-\!\omega_{m}^{2}\right]^{2}
\!+\!\nu^{2}\omega^{2}}\frac{\cos(m\pi x/L_x)}{2}. \label{AvCos}
\end{align}
Eqs.\ (\ref{EqSlow}) and (\ref{EqFast}) have to be supplemented with
the boundary conditions. At finite $q$, coupling to the external
fields is negligible and with high accuracy we can use simple
nonradiative boundary conditions
\begin{equation}
\frac{\partial\tilde{\vartheta}_{\mathbf{k},\beta}}{\partial
x}=0\text{ for }x=0,L_x.
\end{equation}
The stability analysis reduces to computing spectrum of complex
eigenfrequencies $\Omega(\mathbf{k})$ from equations Eqs.\
(\ref{EqSlow}) and (\ref{EqFast}) with nonradiative boundary
conditions and finding out if there are regions in $k$-space where
$\mathrm{Im}[\Omega(\mathbf{k})]>0$.

\subsection{Case of homogenous state in external magnetic field}

\begin{figure}[ptb]
\begin{center}
\includegraphics[width=3.in]{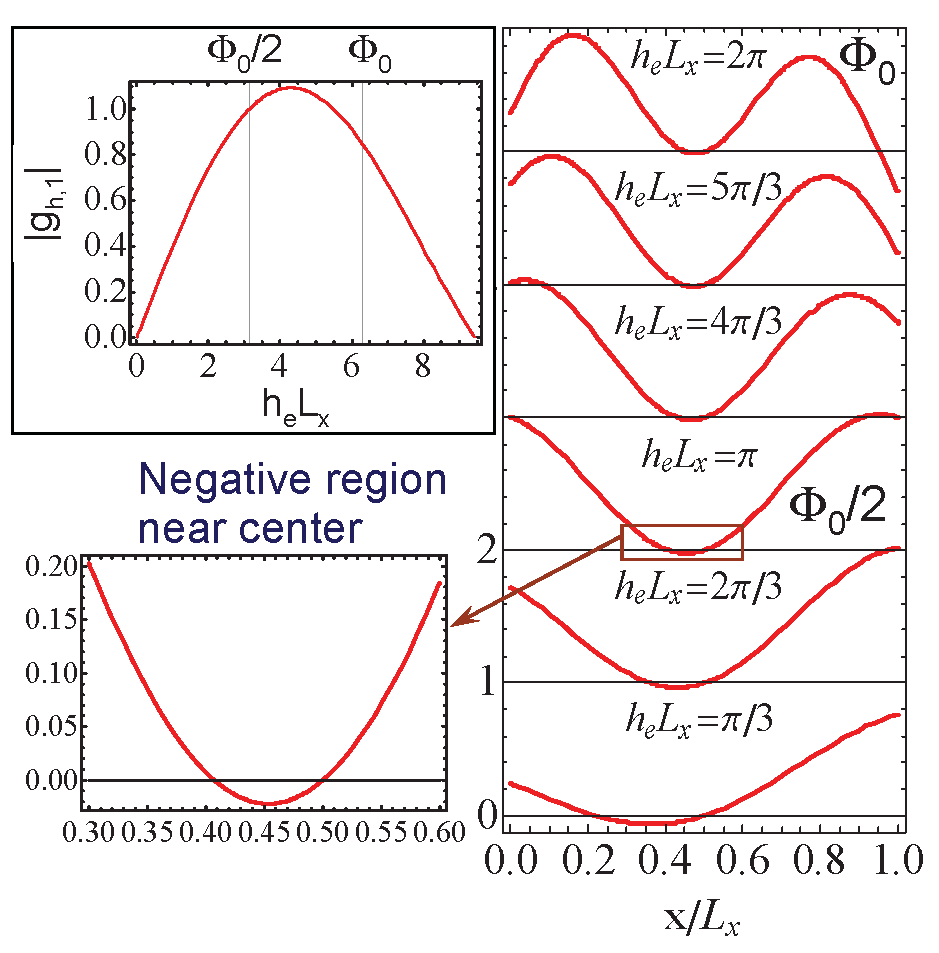}
\end{center}
\caption{(color online) \emph{Upper left} plot shows the field
dependence of the coupling parameter $|g_{h,1}|$, Eq.\
(\ref{FieldCoupl}). \emph{Right} plot illustrates shapes of the
local averaged Josephson coupling $U(x)/U_{\omega}$, Eq.\
(\ref{Ux}), for different magnetic fields and for $\nu
\omega/(\omega_1^2-\omega^2)=0.3$.  \emph{Lower left} plot shows
blowup of the region near the center for $h_eL_x=\pi$ to illustrate
existence of the region with $U(x)<0$ which may lead to the
short-scale instability.} \label{Fig-FieldPlots}
\end{figure}
The above derivation can be directly extended to the case of
the homogeneous state in external magnetic field. Perturbing the
homogeneous solution, $\varphi
_{n}(x,\tau)=\omega\tau+h_{e}x+\theta(x,\tau)+\vartheta_{n}(\mathbf{r},\tau)$,
we obtain Eqs. (\ref{EqSmallDevPh}) and (\ref{EqSmallDevH}) for
small perturbation $\vartheta_{n}(x,\tau)$, where now we have
$g(x)=1$ and $C(x,\tau)\equiv\cos\left(
\omega\tau+h_{e}x+\theta(x,\tau)\right)  $. Separating again the
slow and fast components in the oscillating phase (\ref{PertPhase}),
and using
\begin{align*}
&  C(x,\tau)\vartheta_{n}(x,\tau)\!\approx\!\frac{1}{2}\sum_{q}\cos
[q(n\!+\!1/2)]\exp\left(  -i\Omega\tau\right)  \!\\
&  \times\sum_{\beta=\pm1}\left[  \tilde{\vartheta}_{\mathbf{k},0}\exp\left[
i\beta\left(  \omega\tau+h_{e}x\right)  \right]  \!+\!\tilde{\vartheta
}_{\mathbf{k},\beta}\exp\left(  -i\beta h_{e}x\right)  \right],
\end{align*}
we obtain coupled equations for the slow and fast phase
oscillations, which are identical to Eqs.\ (\ref{EqSlow}) and
(\ref{EqFast}) with replacements
\begin{align*}
\frac{g(x)}{2}\sum_{\beta=\pm1}\tilde{\vartheta}_{\mathbf{k},\beta}
&\rightarrow \frac{1}{2}\sum_{\beta=\pm1}\tilde{\vartheta}_{\mathbf{k},\beta
}\exp\left(  -i\beta h_{e}x\right),\\
\frac{g(x)\tilde{\vartheta}_{\mathbf{k},0}}{2} &\rightarrow
\frac{\tilde{\vartheta}_{\mathbf{k},0}}{2}\exp\left[  i\beta
h_{e}x\right]
\end{align*}
in the righthand sides of these equations.
The average in time cosine $\bar{C}(x)\equiv\!\langle\cos
\varphi\rangle_{\tau}$ can be evaluated as
\[
\bar{C}(x)\!=-\operatorname{Re}\left[  \frac{g_{h,m}\exp\left[
ih_{e} x\right]
}{(1+\alpha_{r})\omega^{2}-\omega_{m}^{2}+i\nu\omega}\right]
\frac{\cos(m\pi x/L_x)}{2}.
\]
In the following sections we will analyze different instabilities of the dynamic coherent states.

\section{Short wave-length instability\label{Sec:ShortWave}}

\subsection{Modulated junction}

Consider first the region $\ell\tilde{q}\gg1$. In this case instability
develops only for the homogeneous in the $y$-direction perturbations,
$k_{y}=0$, and we only consider such perturbations. In this regime the
derivative term in the fast part (\ref{EqFast}) becomes small and can be
neglected giving the estimate
\[
\tilde{\vartheta}_{\mathbf{k},\beta}\approx\frac{g(x)}{2}\left(  \frac{\left(
\Omega-\beta\omega\right)  ^{2}}{1+\alpha\tilde{q}^{2}}+i\nu_{c}\left(
\Omega-\beta\omega\right)  \right)  ^{-1}\tilde{\vartheta}_{\mathbf{k},0},
\]
meaning that, roughly,
$\tilde{\vartheta}_{\mathbf{k},\beta}\sim\bar{\vartheta}
_{\mathbf{k},0}/\omega^{2}\ll\bar{\vartheta}_{\mathbf{k},0}$.
Substituting this estimate into the equation for the slow part
(\ref{EqSlow}), we conclude that coupling to the fast terms in this
regime is weak and can be neglected. Therefore, the equation for the
slow part becomes
\begin{equation}
\left[  \tilde{\Omega}^{2}-U(x)\right]
\tilde{\vartheta}_{\mathbf{k}
,0}+G_{q,0}^{-2}\frac{\partial^{2}\tilde{\vartheta}_{\mathbf{k},0}}{\partial
x^{2}}=0, \label{HighqEq}
\end{equation}
where $U(x)=g(x)\bar{C}(x)$ determines the local plasma frequency. For the
lowest mode we obtain
\begin{equation}
U(x)=\frac{g(x)g_{1}\left[  \omega_{1}^{2}-(1+\alpha_{r})\omega^{2}\right]
/2}{\left[  \omega_{1}^{2}-(1+\alpha_{r})\omega^{2}\right]  ^{2}+\nu^{2}
\omega^{2}}\cos(\pi x/L_{x}). \label{EffPotential}
\end{equation}
For large $q$ we can use the simple boundary condition
$\partial\tilde{\vartheta }_{\mathbf{k},0}/\partial x=0$ at the
edges, $x=0,\ L_{x}$. As the local \textquotedblleft spring
constant\textquotedblright\ $U(x)$ determines the local plasmon
frequency $\Omega_{\mathrm{loc}}\propto\pm\sqrt{U(x)}$, the
existence of regions with negative $U(x)$ is a potential source of
instability. This instability may be eliminated by the gradient term
which we have to verify. Consider for definiteness the lowest mode
and \emph{negative} coupling parameter, $g_{1}<0$. The simplest
modulation giving such coupling is monotonically increasing $g(x)$.
In this case at $\omega<\omega_{1}$ the \textquotedblleft spring
constant\textquotedblright\ $U(x)$ is negative at $x<L_{x}/2$ and
has minimum near the edge $x=0$ where $g(x)$ is the smallest.
Therefore, the region for potential instability is located near the
left edge, at $x\ll L_{x}$. As the most unstable mode is localized
near $x=0$, we can expand $U(x)$ with respect to $x$
\begin{align*}
&  g(x)\cos(\pi x/L_{x})\approx g(0)\left[  1+a_{1}x-\frac{a_{2}}{2}
x^{2}\right] \\
&  \text{with }a_{1}=g^{\prime}(0)/g(0),\ a_{2}=\left(  \pi/L_{x}\right)
^{2}-g^{\prime\prime}(0)/g(0)
\end{align*}
giving the equation
\begin{align*}
&  \left[  \tilde{\Omega}^{2}+U_{\omega}\left(  1+a_{1}x-\frac{a_{2}}{2}
x^{2}\right)  \right]  \tilde{\vartheta}_{\mathbf{k},0}+G_{q}^{-2}
\frac{\partial^{2}\tilde{\vartheta}_{\mathbf{k},0}}{\partial x^{2}}=0\\
&  \text{with }U_{\omega}\equiv\frac{g(0)|g_{1}|\left[  \omega_{1}
^{2}-(1+\alpha_{r})\omega^{2}\right]  /2}{\left[
\omega_{1}^{2}-(1+\alpha _{r})\omega^{2}\right]
^{2}+\nu^{2}\omega^{2}}>0.
\end{align*}
From this equation we estimate,
\[
\tilde{\Omega}^{2}\approx-U_{\omega}\left[
1+\frac{a_{1}^{2}}{2a_{2}}
-C\sqrt{\frac{a_{2}}{U_{\omega}G_{q}^{2}}}\right]
\]
with
\[
\frac{a_{2}}{U_{\omega}G_{q}^{2}}\!=\!\frac{\left(  \pi/L\right)
^{2}-g^{\prime\prime}(0)/g(0)}{1+\ell^{2}\tilde{q}^{2}/[1\!-\!i\Omega\nu
_{ab}]}\frac{\left[  \omega_{1}^{2}\!-\!(1\!+\!\alpha_{r})\omega^{2}\right]
^{2}\!+\!\nu^{2}\omega^{2}}{g(0)|g_{1}|\left[  \omega_{1}^{2}
\!-\!(1\!+\!\alpha_{r})\omega^{2}\right]  /2}
\]
and $C\sim1$. Roughly, the system can only be stable if
$|a_{2}U_{\omega} G_{q}^{-2}|\gtrsim1$ for all $q$'s. As
$|G_{q}^{-2}|$ has minimum at $q=\pi$ ($\tilde{q}=2$), it is
sufficient to check the stability condition at this value of $q$.
Skipping numerical factors on the order of unity, we can
approximately write the stability criterion as
\begin{equation}
\frac{g(0)g_{1}\left(  \omega_{1}^{2}-\omega^{2}\right)  }{\left(
\omega _{1}^{2}-\omega^{2}\right)
^{2}+\nu^{2}\omega^{2}}<\frac{\pi^{2}}{4\ell ^{2}L_{x}^{2}}.
\label{StabCondHighq}
\end{equation}
This condition is very hard to satisfy because the right-hand side
of this equation is very small due to $\ell L_{x}\gg1$. Therefore,
\emph{the homogeneous solution is almost always unstable with
respect to alternating deformations localized near the edge with
suppressed Josephson coupling}. The formal reason for the large-$q$
instability is that the local spring constant\ (\ref{EffPotential})
changes sign due to the factor $\cos(\pi x/L_{x})$. This instability
would be completely eliminated if the modulation function $g(x)$
would also changes sign in the middle. It is practically impossible
to prepare such modulation artificially. The best artificial
modulation for which the homogeneous state is stable with respect to
the short-scale perturbations is steplike modulation for which the
Josephson current \emph{completely suppressed} in one half of the
stack, $g(x)=0$ for $x<L_{x}/2$. In this case the plasma frequency
is zero in this half and stability is achieved due to the gradient
term in Eq.\ (\ref{HighqEq}).

We will demonstrate below that the alternating-kink solution is
stable with respect to the large-$q$ perturbation because it
generates an effective modulation changing sign in the middle of the
stack making the spring constant positive in the whole stack.
Another interesting case is the stack in the magnetic field
corresponding to half flux
quantum per junction. In this case the linearly growing contribution to the phase
changes from $0$ to $\pi$ across the junction which is similar to
a sign-changing modulation. We will consider these cases in more
detail in the following subsections.

\subsection{Absence of short wave-length instability for the alternating-kink
state}

The stability analysis for the modulated system can be directly applied to
alternating-kink state. It is crucial, however, that the effective modulation
function $g_{k}(x)$ is close to a step function changing from $1$ to $-1$ in
the middle of the stack, i.e., it changes sign in the middle which compensates
change of sign in the factor $\cos(\pi x/L_{x})$. This eliminates the
short-scale instability. Indeed, the effective \textquotedblleft spring
constant\textquotedblright\ (\ref{EffPotential}), which determines the local
oscillation frequency, can be evaluated as
\begin{align}
U_{k}(x)  &  =g_{k}(x)\bar{C}(x)\nonumber\\
&  \approx\frac{g_{k,1}\left[
\omega_{1}^{2}-(1+\alpha_{r})\omega^{2}\right] /2}{\left[
\omega_{1}^{2}-(1+\alpha_{r})\omega^{2}\right]  ^{2}+\nu^{2}
\omega^{2}}|\cos(\pi x/L_{x})|.
\end{align}
It only touches zero in the midpoint and never goes negative within the stack.
Therefore, in contrast to the homogeneous state in modulated junction,
\emph{the alternating-kink state is stable with respect to short-scale
perturbations}.

\subsection{Test of short-scale instability in modulated stack with numerical simulations}

\begin{figure}[ptb]
\begin{center}
\includegraphics[width=3.4in]{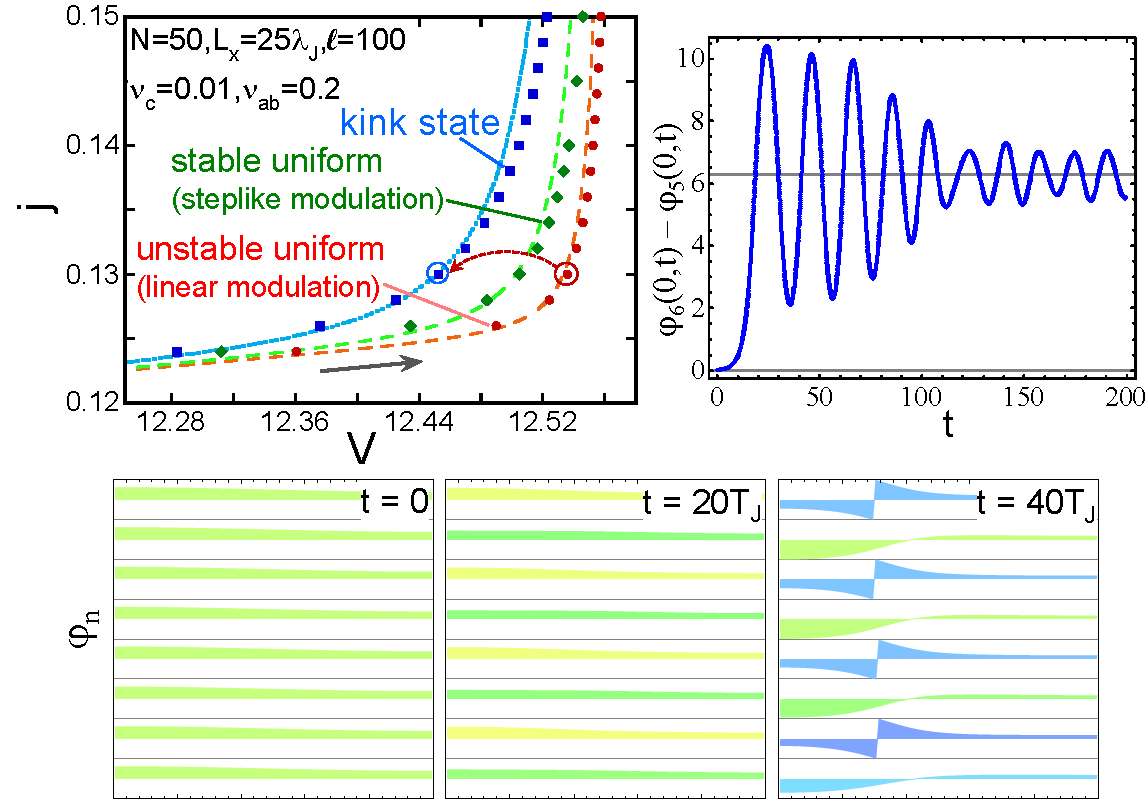}
\end{center}
\caption{(color online)\emph{Upper left} plot shows current-voltage
dependences near the resonance for three dynamic states: (i)
unstable uniform state in the stack with linear modulation of the
Josephson current, $j_J(x)=j_{J0}(1+r(2x/L-1))$ with $r=0.4$, (ii)
kink state, and (iii) stable uniform state in the stack with
steplike modulation of the Josephson current,
$j_J(x)=j_{J0}\Theta(2x/L-1)$. Dashed lines show corresponding
theoretical curves. Instability for the first state was triggered at
the point $j=0.13$ marked at the plot. \emph{Lower plots} shows
snapshots of the phase distributions in the eight bottom junctions
for $t=0$, $20T_J$, and $40T_J$, where $T_J=2\pi/\omega_J \approx
0.5$ is the period of Josephson oscillations for the uniform state.
All phases are shifted to the range $(-\pi,\pi)$. The snapshots
illustrate development of instability near the left edge.
\emph{Upper right} plot shows the time evolutions of the difference
between phases in the 6th and 5th junctions at the left edge.  This
difference changes from $0$ corresponding to the uniform state to
$2\pi$ corresponding to the kink state.}
\label{Fig-SimShortScaleInst}
\end{figure}

We verified the short scale instability in modulated junctions using
numerical simulations described in Appendix \ref{App-Num}. First, we
probed the stability of the uniform state in the stack with linear
modulation of the Josephson current, $j_J(x)=j_{J0}(1+r(2x/L-1))$.
We used the modulation parameter $r=0.4$.  If we start from the
uniform $n$-independent state and solve the dynamic equations
without noise than instability does not develop and we can trace the
current-voltage dependence corresponding to this uniform state, see
Fig.\ \ref{Fig-SimShortScaleInst}. However, if we add to the phases
a small alternating perturbations, $(-1)^n\delta\varphi$, than we
observe that the uniform state blows up and, after extended time
evolution, it converges to the dynamic kink state. The initial stage
of this time evolution is illustrated in the lower part of Fig.\
\ref{Fig-SimShortScaleInst}. From the phase snapshots we can see
that the instability develops near the edge with suppressed
Josephson current, as the analytical analysis predicts. The upper
right plot in Fig.\ \ref{Fig-SimShortScaleInst} shows the time
evolutions of the difference between phases in the 6th and 5th
junctions at the left edge.  We can see that this difference evolves
from $0$ corresponding to the uniform state to the value $2\pi$
corresponding to the kink state. This transition is accompanied by
large-amplitude oscillations with the period much longer than the
period of Josephson oscillations. These oscillations are a
consequence of the small c-axis dissipation parameter, $\nu_c=0.01$,
which we used in our calculations.

We also verified that the uniform state remains stable in the stack
with steplike modulation of the Josephson current, $j_J(x)=j_{J0}$
for $x>L_{x}/2$ and $j_J(x)=0$ for $x<L_{x}/2$. The current-voltage
dependence for this state is also shown in Fig.\
\ref{Fig-SimShortScaleInst}.

\subsection{Homogeneous state in magnetic field}

\begin{figure}[ptb]
\begin{center}
\includegraphics[width=2.45in]{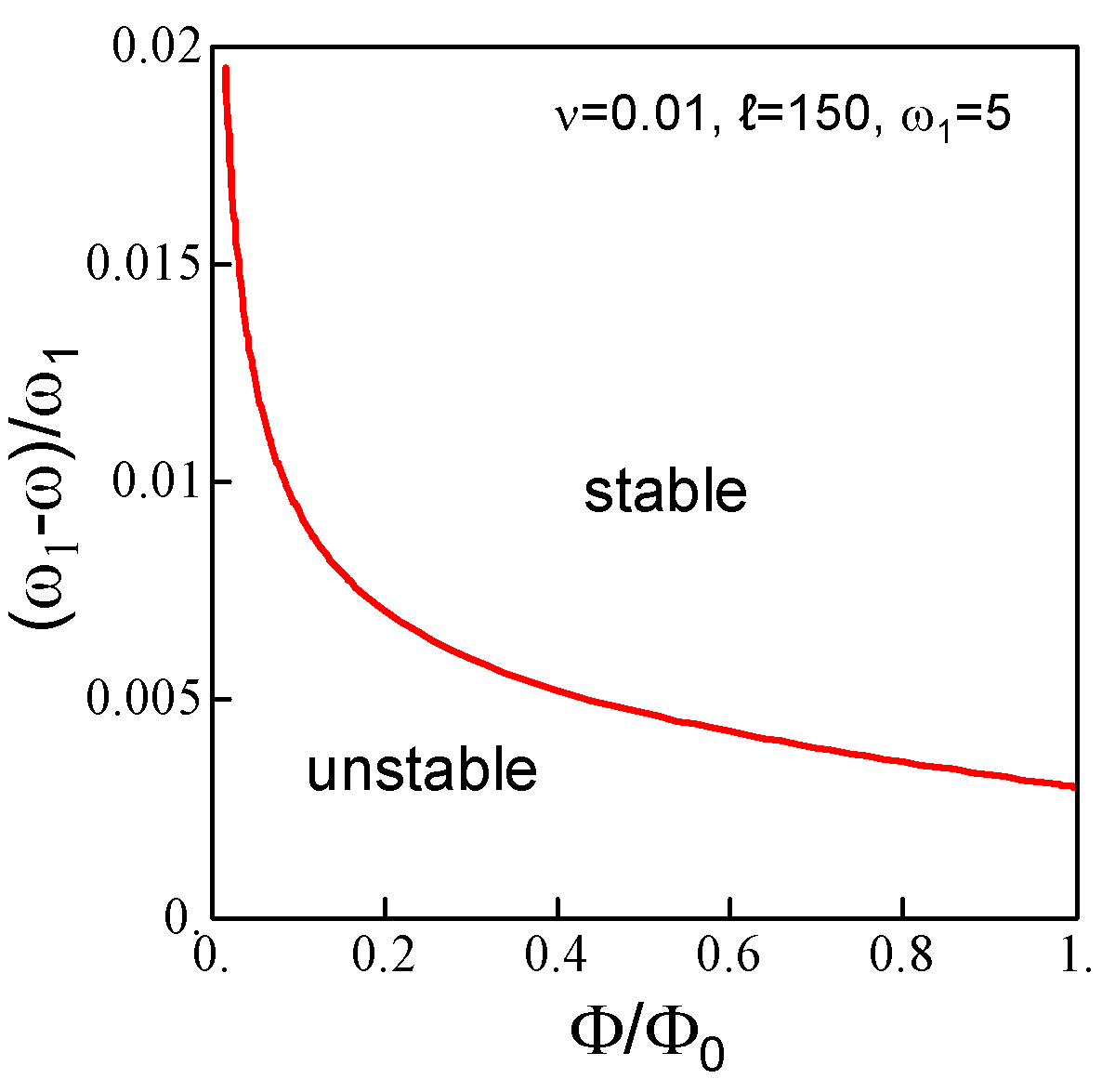}
\end{center}
\caption{(color online) Stability range of the homogeneous state in magnetic
field with respect to the short-scale phase deformations based on
Eq.\ (\ref{StabCritHighqH}) with the listed representative parameters.}
\label{Fig-ShrtScaleH}
\end{figure}
Due to the complex coupling function in the case of finite magnetic
field, the short-range stability has features which are special for
this case. In the regime $\ell\tilde{q}\gg1$ the coupling to the
fast phase can be neglected, as for the modulated stack case. This
means that the slow part again obeys Eq.\ (\ref{HighqEq}) in which
the local \textquotedblleft spring constant\textquotedblright\
$U(x)$ now is simply given by the average cosine, $U(x)=\bar{C}(x)$.
For the fundamental mode, we obtain
\begin{align}
U(x)  &  =U_{\omega} \left\{
-\frac{\nu\omega}{\omega_{1}^{2}-\omega^{2}} \cos\left[ h_{e}\left(
x-\frac{L_x}{2}\right)  \right]
\right.\nonumber\\
& -\left.  \sin\left[  h_{e}\left(  x-\frac{L_x}{2}\right)  \right]
\right\} \cos\left(\frac{\pi x}{L_x}\right),\label{Ux}
\\
\text{with  }U_{\omega}  & =\frac{2h_{e}L_x\cos\left[
h_{e}L_x/2\right] }{\pi^{2}-\left(  h_{e}L_x\right)
^{2}}\frac{\omega_{1}^{2}-\omega^{2}}{\left(
\omega^{2}-\omega_{1}^{2}\right)  ^{2}+\nu^{2}\omega^{2}}.\nonumber
\end{align}
For simplicity, we omit here the factor $(1+\alpha_{r})$ in front of
$\omega^{2}$ which has very little influence on the short
wave-length stability. Shapes of this function at different fields
is illustrated in the upper right plot of Fig.\ \ref{Fig-FieldPlots}
for $\nu \omega/(\omega_1^2-\omega^2)=0.3$. Note that $U(x)$ always
changes sign at the center, $x=L_x/2$, see, e.g., lower left plot in
Fig.\ \ref{Fig-FieldPlots}. However, for $\nu\omega\ll
\omega_{1}^{2}-\omega^{2}  $ the region of negative $U(x)$ is very
narrow and its existence does not automatically imply instability.
We analyze the central region as the most prone to instability. For
$\omega_{1}^{2}-\omega^{2} \gg\nu\omega$ the \textquotedblleft
spring constant\textquotedblright\ behaves near $x=L_x/2$ as
$U(x)\approx U_{\omega}\left[\nu\omega/\left(
\omega_{1}^{2}\!-\!\omega^{2}\right)+h_{e}\tilde{x}  \right]
\pi\tilde{x}/L_x$ with $\tilde {x}=x-L_x/2$, and Eq. (\ref{HighqEq})
becomes
\[
\left[  \tilde{\Omega}^{2}-U_{\omega}\left(
\frac{\nu\omega}{\omega_{1}
^{2}\!-\!\omega^{2}}\!+\!h_{e}\tilde{x}\right)
\frac{\pi\tilde{x}}{L_x}\right]
\tilde{\vartheta}_{\mathbf{k},0}+G_{q}^{-2}\frac{\partial^{2}\tilde{\vartheta}_{\mathbf{k},0}}
{\partial\tilde{x}^{2}}\!=\!0.
\]
From this linear-oscillator-type equation, we derive equation for the complex
eigenfrequency
\begin{align*}
&  \Omega^{2}/(1+\alpha\tilde{q}^{2})+i\nu_{c}\Omega\\
&  =U_{\omega}\left[  -\frac{\nu^{2}\omega^{2}\pi/L_x}{4h_{e}\left(
\omega _{1}^{2}-\omega^{2}\right)  ^{2}}+\sqrt{\frac{\pi
h_{e}/L_x}{G_{q}^{2}U_{\omega }}}\right],
\end{align*}
which gives the stability criterion
\[
\sqrt{\frac{\pi h_{e}/L_x}{2\ell^{2}(1-\cos
q)U_{\omega}}}>\frac{\nu^{2} \omega^{2}\pi/L_x}{4h_{e}\left(
\omega_{1}^{2}-\omega^{2}\right)^{2}}.
\]
For the most "dangerous" mode at $q=\pi$, assuming $\omega_{1}^{2}-\omega
^{2}\gg\nu\omega$, we obtain the following stability criterion \
\begin{equation}
\omega_{1}^{2}-\omega^{2}>\left[
\frac{\pi|g_{h,1}|\ell^{2}\nu^{4}\omega^{4} }{8h_{e}^{3}L_x}\right]
^{1/5}. \label{StabCritHighqH}
\end{equation}
Large value of $\ell^{2}$ in the righthand side is compensated by
small value of $\nu^{4}\omega^{4}$. Note that increasing dissipation
reduces the stability range. Taking typical values $h_{e}\approx5$,
$\omega_{1}\approx5$, $h_{e}L=\pi$, $\ell=150$, and $\nu=0.002$,
this inequality gives $\left( \omega_{1}-\omega\right)
/\omega_{1}>1.3\cdot10^{-3}$ and for larger dissipation, $\nu=0.01$,
$(\omega_{1}-\omega)/\omega_{1}>0.005$. This conditions are not too
restrictive. A representative stability region in the magnetic
field is illustrated in Fig.\ \ref{Fig-ShrtScaleH}. We conclude that
the homogeneous state in the magnetic field corresponding to half
flux quantum per junction remains stable with respect to the
short-scale deformations if the frequency is not too close to the
resonance.

\section{Long-wave length stability\label{Sec:LongWave}}

 We analyze now the acoustic type instability at
 very small $k_{y}$ and $q$. With very minor
 modifications, this analysis applies to all dynamic
 states considered in this paper. Obviously, it is
 most relevant for the systems which are stable with
 respect to the short-length deformations. Consider
 equation (\ref{EqFast}) for the fast components
 $\tilde{\vartheta}_{\mathbf{k},\pm1}$. At small
 $k_{y}$ and $q$ this equation gives the resonance
 solution. To obtain approximate solution for
 $\tilde{\vartheta}_{\mathbf{k},\pm1}$, we will keep
 only the resonance term,
 $\tilde{\vartheta}_{\mathbf{k},\beta}(x)\approx\psi_{\mathbf{k},\beta}
 \cos(m\pi x/L_{x})$. Further analysis shows that
 the coordinate dependence of
 $\tilde{\vartheta}_{\mathbf{k},0}$ is weak and can
 be neglected. In this case, the mode amplitude
 $\psi_{\mathbf{k},\beta}$ can be found following
 the same reasoning as for the homogeneous solution
 leading to
\[
\psi_{\mathbf{k},\beta}\approx\frac{g_{m}\tilde{\vartheta}_{\mathbf{k},0}
/2}{\tilde{\Omega}_{\beta}^{2}-G_{q,\beta}^{-2}\left(
\omega_{m}^{2} +k_{y}^{2}\right)  }.
\]
Using this result, we present Eq.\ (\ref{EqSlow}) for the slow
component $\tilde{\vartheta}_{\mathbf{k},0}$ as
\begin{align}
&  \left[  \tilde{\Omega}_{0}^{2}-g(x)\bar{C}(x)-G_{q,0}^{-2}k_{y}^{2}\right]
\tilde{\vartheta}_{\mathbf{k},0}+G_{q,0}^{-2}\frac{\partial^{2}\tilde
{\vartheta}_{\mathbf{k},0}}{\partial x^{2}}\nonumber\\
&
=\frac{g(x)}{4}\sum_{\beta=\pm1}\frac{g_{m}\tilde{\vartheta}_{\mathbf{k}
,0}\cos(m\pi
x/L_{x})}{\tilde{\Omega}_{\beta}^{2}-G_{q,\beta}^{-2}\left(
\omega_{m}^{2}+k_{y}^{2}\right) }. \label{SlowCompEq}
\end{align}
The typical length scale for its variation, $l_{\Omega}=\left(
\sqrt {\left\vert
\tilde{\Omega}_{0}^{2}-G_{q,0}^{-2}k_{y}^{2}\right\vert }
|G_{q,0}|\right)  ^{-1}$ exceeds the stack width $L_{x}$ because we
consider the case of small $k_{y}$, $q$, and $\Omega$ when
$G_{q,0}\sim1$ and $\Omega,k_{y}\ll1/L_{x}$. This allows us to
neglect $x$ dependence of $\tilde{\vartheta}_{\mathbf{k},0}$. In
this case $g(x)\bar{C}(x)$ and $g(x)\cos(m\pi x/L_{x})$ can be
replaced by their averages over $x$
\begin{align*}
&  \left\langle g(x)\bar{C}(x)\!\right\rangle
_{x}\approx-\frac{g_{m} ^{2}\left[
(1+\alpha_{r})\omega^{2}-\omega_{m}^{2}\right]  /4}{\left[
(1+\alpha_{r})\omega^{2}-\omega_{m}^{2}\right]  ^{2}+\left(
\nu_{c}+\nu
_{r}\right)  ^{2}\omega^{2}},\\
&  \left\langle g(x)\cos(m\pi x/L_{x})\right\rangle _{x}=g_{m}/2.
\end{align*}
We also can neglect the charging effects because at the typical wave vector
$q\sim\pi/N$ and $\alpha\lesssim0.1$ the charging correction $\alpha q^{2}$ is
tiny and has only minor influence on stability criteria. In these
approximations, the equation for $\Omega(q)$ becomes
\begin{equation}
\Omega^{2}-\frac{k_{y}^{2}}{1+k_{z}^{2}/[1\!-\!i\Omega\nu_{ab}]}+i\nu
_{c}\Omega=\frac{g_{m}^{2}}{8\omega_{m}}\mathcal{G}(\Omega,\mathbf{k}
)\label{FreqEq}
\end{equation}
with $k_{z}=2\ell\sin(q/2)\approx\ell q$, $\mathbf{k}=(k_{y},k_{z})$ and
\begin{align*}
&  \mathcal{G}(\Omega,\mathbf{k})=-\frac{2\omega_{m}\left[  (1+\alpha
_{r})\omega^{2}-\omega_{m}^{2}\right]  }{\left[  (1+\alpha_{r})\omega
^{2}-\omega_{m}^{2}\right]  ^{2}+\nu^{2}\omega^{2}}\\
&  +\!\sum_{\beta=\pm1}\!\omega_{m}\!\left[  \left(
\Omega\!-\!\beta \omega\right)
^{2}\!-\!\frac{\omega_{m}^{2}+k_{y}^{2}}{1\!+\!\frac{k_{z}^{2}
}{1\!-\!i\nu_{ab}\left(  \Omega\!-\!\beta\omega\right)
}}\!+\!i\nu_{c}\left( \Omega\!-\!\beta\omega\right)  \right]  ^{-1}.
\end{align*}
If we eliminate the radiaton corrections, $\nu_{r}$ and
$\alpha_{r}$, which only appear at very small $k_{y}$ and $k_{z}$
then this function satisfies the translational invariance condition
$\mathcal{G}(0,0)=0$. Note that we can not take true limit
$k_{y},k_{z}\rightarrow0$, because we consider a finite-size stack
with geometrical sizes smaller than the wave length of outside
radiation.

As the Josephson frequency $\omega$ is close to the cavity-mode
frequency $\omega_m$, we can keep only the dominating resonance
terms. To simplify presentation of the second term, we introduce the
shift of the plasma frequency at finite $k_{y}$ and $k_{z}$ with
respect to the homogeneous mode,
$\Delta_{\mathbf{k}}=\omega_{p}(k_{m},0,0)-\omega_{p}(k_{m},k_{y},k_{z})$,
and mode damping parameter, $\nu_{\mathbf{k}}$,
\begin{align*}
\Delta_{\mathbf{k}}+i\frac{\nu_{\mathbf{k}}}{2}  &
=\frac{1}{2\omega_{m} }\left(
\omega_{m}^{2}-\frac{\omega_{m}^{2}+k_{y}^{2}}{1+k_{z}^{2}/\left(
1-i\nu_{ab}\left(  \Omega-\beta\omega\right)  \right)  }\right) \\
&  \approx\frac{\omega_{m}}{2}\left(
\frac{k_{z}^{2}}{1+\nu_{ab}^{2}
\omega_{m}^{2}}-\frac{k_{y}^{2}}{k_{m}^{2}}\right)
+i\frac{\omega_{m}^{2}}
{2}\frac{\nu_{ab}k_{z}^{2}}{1+\nu_{ab}^{2}\omega_{m}^{2}}.
\end{align*}
Introducing also the resonance detuning $\delta_{\omega}=\omega-\omega_{m}$,
we simplify $\mathcal{G}(\Omega,\mathbf{k})$ as
\begin{align}
\mathcal{G}(\Omega,\mathbf{k})  &  \approx-\frac{\delta_{\omega}+\delta_{r}
}{\left(  \delta_{\omega}+\delta_{r}\right)  ^{2}+\nu^{2}/4}\nonumber\\
&  +\frac{\delta_{\omega}+\Delta_{\mathbf{k}}}{\left(
\delta_{\omega} +\Delta_{\mathbf{k}}\right)  ^{2}-\left(
\Omega+i\nu_{k}/2\right)  ^{2}}
\label{ResonDisp}
\end{align}
with $\delta_{r}=\alpha_{r}\omega_{m}/2$ being the radiation shift of the
resonance frequency and
\begin{equation}
\nu_{k}\approx\nu_{c}+\frac{\omega_{m}^{2}\nu_{ab}k_{z}^{2}}{1+\nu_{ab}
^{2}\omega_{m}^{2}} \label{q-dissip}
\end{equation}
is the damping of the plasma mode at finite wave vector (we neglect
small terms on the order of $k_{z}^{2}k_{y}^{2}$).

Due to the resonance structure of $\mathcal{G}(\Omega,\mathbf{k})$, in the
limit $\Delta_{\mathbf{k}},|\delta_{\omega}|\ll g_{m}/\sqrt{8\omega_{m}}$ it
typically exceeds the righthand side of Eq. (\ref{FreqEq}) and the dispersion
equation is approximately given by $\mathcal{G}(\Omega,\mathbf{k})=0$. In this
approximation we obtain the following result for the eigenfrequencies
\begin{equation}
\Omega_{\pm}(\mathbf{k})\approx-i\frac{\nu_{k}}{2}\pm\!\sqrt{\left(
\tilde{\delta}_{\omega}\!+\!\tilde{\Delta}_{\mathbf{k}}\right)  \left(
\tilde{\Delta}_{\mathbf{k}}\!-\!\frac{\nu^{2}}{4\tilde{\delta}_{\omega}
}\right)  },\label{EigenFreqLong}
\end{equation}
where we introduced new notations
$\tilde{\delta}_{\omega}=\delta_{\omega }+\delta_{r}$ and
$\tilde{\Delta}_{\mathbf{k}}=\Delta_{\mathbf{k}}-\delta_{r}$ to
absorb the radiation frequency shift. Only the mode $\Omega_{+}
(\mathbf{k})$ is potentially unstable. The instability takes place when
the expression under the square root is negative and the imaginary
square root exceeds the first term. This leads to the following
condition for the instability in the $(k_{y},k_{z})$ plane
\begin{equation}
\left(  \tilde{\delta}_{\omega}\!+\!\tilde{\Delta}_{\mathbf{k}}\right)
\!\left(  \!\frac{\nu^{2}}{4\tilde{\delta}_{\omega}}\!-\!\tilde{\Delta
}_{\mathbf{k}}\right)  \!>\!\frac{\nu_{k}^{2}}{4}.\label{InstabCrit}
\end{equation}
Analysis of this criterion shows that the long-range stability is determined
by several factors including behavior of the plasmon frequency shift
$\Delta_{\mathbf{k}}$ and the relation between the damping of the homogeneous mode
$\nu$ and damping of excited modes at finite wave vectors, $\nu_{k}$. As the
shift $\tilde{\Delta}_{\mathbf{k}}$ may take both positive and negative values
depending on $k_{y}$ and $k_{z}$, it is more transparent to find the instability
range for this parameter,
\begin{equation}
\left\vert
\tilde{\Delta}_{\mathbf{k}}+\frac{\tilde{\delta}_{\omega}}{2}
-\frac{\nu^{2}}{8\tilde{\delta}_{\omega}}\right\vert
<\frac{1}{2}\sqrt{\left(
\tilde{\delta}_{\omega}+\frac{\nu^{2}}{4\tilde{\delta}_{\omega}}\right)
^{2}-\nu_{k}^{2}}\label{InstDeltak}
\end{equation}
In particular, there is no instability in the $k_{z}$-region satisfying the
condition
\[
\nu_{k}>|\tilde{\delta}_{\omega}|+\frac{\nu^{2}}{4|\tilde{\delta}_{\omega}|}
\]
which explicitly can be written as condition for $k_{z}$
\begin{align}
k_{z} &  >k_{z,i},\label{NoInstab-q}\\
k_{z,i}^{2} &
=\frac{1+\nu_{ab}^{2}\omega_{m}^{2}}{\nu_{ab}\omega_{m}^{2} }\left[
|\tilde{\delta}_{\omega}|+\frac{(\nu_{c}+\nu_{r})^{2}}{4|\tilde
{\delta}_{\omega}|}-\nu_{c}\right]  .\nonumber
\end{align}
Qualitatively, we can conclude that the in-plane dissipation
suppresses instability until $\nu_{ab}\omega_{m}<1$, while the
radiation damping enhances the instability. As for a finite-size
stack the $k_{z}$-values are limited from below, $k_{z}>\ell\pi/N$,
the above equation also determines the stable frequency range for a
stack with given size. We can see that this range expands with
decreasing $N$.
\begin{figure}[tb]
\begin{center}
\includegraphics[width=3.4in]{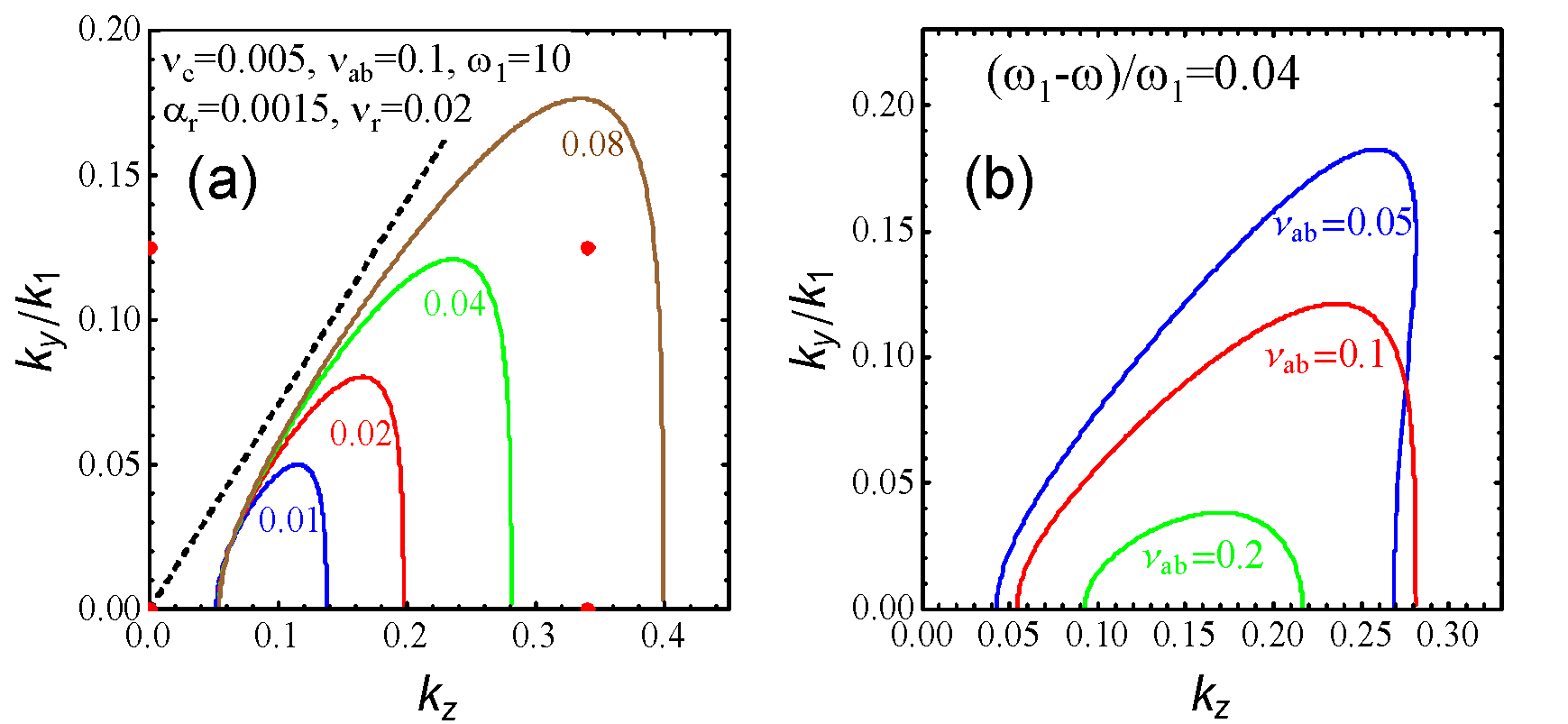}
\end{center}
\caption{(color online)(a) Evolution of the instability region for
the long-wave deformations in the $k_{z}$-$k_{y}$ plane with
approaching the resonance from below. Unstable regions are inside
the domes. Used representative parameters are shown in the plot. The
curves are marked by the value $(\tilde{\omega}
_{1}-\omega)/\omega_{1}=-\tilde{\delta}_{\omega}/\omega_{m}$. Dashed
line shows dependence
$k_{y}/k_{1}=k_{z}/\sqrt{1\!+\!\nu_{ab}^{2}\omega_{1}^{2}}$
corresponding to condition $\omega_{p}(k_{1},k_{y},k_{z})=\omega_{p}
(k_{1},0,0)$. The points illustrate discrete wave vectors for a
finite-size stack. In this example, the system becomes unstable for
$(\tilde{\omega} _{m}-\omega)/\omega_{m}$ between 0.04 and 0.08.
(b)Evolution of the instability region in the $k_{z} $-$k_{y}$ plane
at fixed frequency with increasing the in-plane dissipation
parameter $\nu_{ab}$.} \label{Fig-IntabReg-kykz}
\end{figure}
\begin{figure}[tb]
\begin{center}
\includegraphics[width=3.5in]{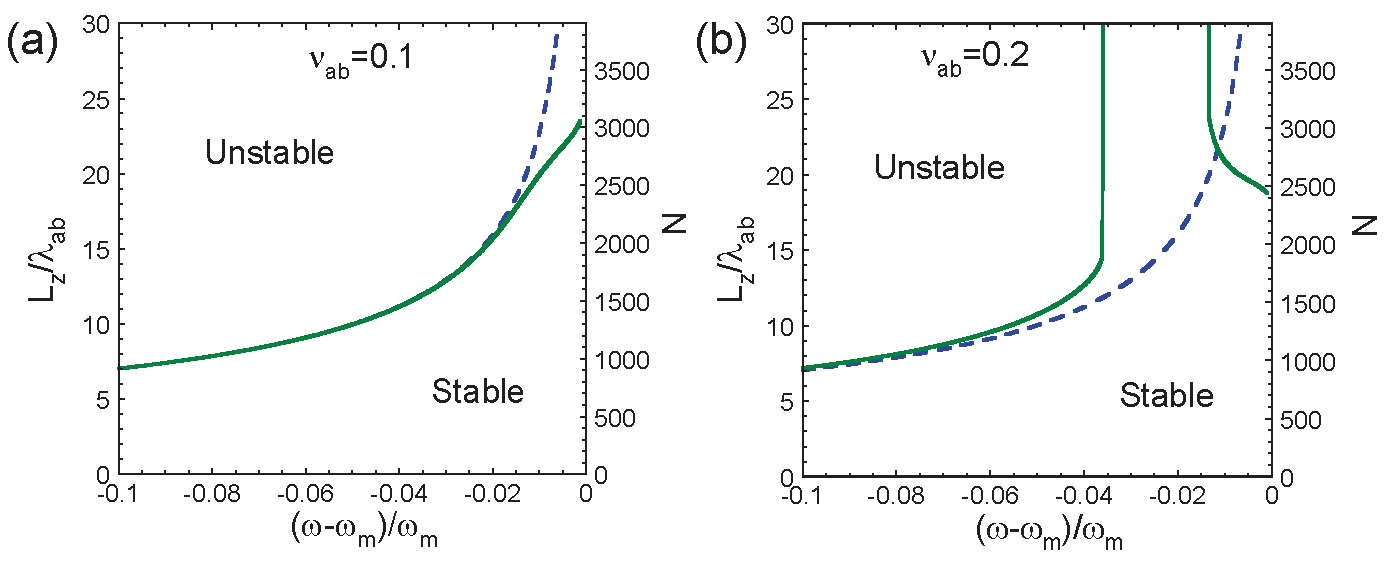}
\end{center}
\caption{(color online) Stability range with respect to the
long-wave deformations near the resonance as a function of the stack
height based on Eq. (\ref{kz0}). Two regimes are illustrated in
plots (a) and (b), which are controlled by the in-plane dissipation.
Dashed lines show stability boundaries without radiation
corrections. In the calculations we used the following
representative parameters, $\nu_{c}=0.005$, $\omega_{1}=10$,
$\ell=130$. For radiation parameters we used $\alpha_{r}=3.3
\cdot10^{-4}L_{z}/\lambda_{ab}$ and $\nu_{r}/\omega_{1}=5.2
\cdot10^{-4}L_{z}/\lambda_{ab}$, where the numerical coefficients
were estimated assuming $\lambda_{ab}=0.25$$\mu$m, $L_{x}=80$$\mu$m,
$\epsilon_{c}=12$, and $\ln(C/k_{\omega}L_{z})=4$.}
\label{Fig-SizeFreqDiag-vab}
\end{figure}

In small-$k_{z}$ range where the condition (\ref{NoInstab-q}) is not satisfied
there may be a range of $k_{y}$ where the system is unstable. We can find from
Eq. (\ref{InstDeltak}) an explicit presentation for this range
\begin{align}
&  \left\vert \frac{k_{y}^{2}}{k_{m}^{2}}-\frac{k_{z}^{2}}{1\!+\!\nu_{ab}
^{2}\omega_{m}^{2}}\!+\!\frac{\tilde{\delta}_{\omega}}{\omega_{m}}
\!-\!\frac{\nu^{2}}{4\tilde{\delta}_{\omega}\omega_{m}}\right\vert \nonumber\\
&  <\sqrt{\left(  \frac{\tilde{\delta}_{\omega}}{\omega_{m}}+\frac{\nu^{2}
}{4\tilde{\delta}_{\omega}\omega_{m}}\right)  ^{2}+\frac{\nu_{k}^{2}}
{\omega_{m}^{2}}}. \label{Instab-kykz}
\end{align}
The instability regions in the $(k_{y},k_{z})$-plane for the
different Josephson frequencies near the resonance are illustrated
in Fig. \ref{Fig-IntabReg-kykz}a. One can see that the instability
region rapidly shrinks with approaching the resonance. Vanishing of
instability at large $k_{z}$ is caused by increasing mode damping
due to the in-plane dissipation. This is illustrated in Fig.\
\ref{Fig-IntabReg-kykz}b where we plot the instability region at
fixed Josephson frequency for different in-plane dissipation
parameter $\nu_{ab}$. We can see that the instability regions
shrinks with increasing $\nu_{ab}$. Existence of the instability
region in the $(k_{y},k_{z})$-plane does not automatically imply
instability in real junction stacks. In a finite-size stack a
discrete set of the wave vectors is allowed, $k_{y,m}
\!=\!m\pi/L_{y}$, $k_{z,n}\!=\!n\ell \pi/N$. The system is only
unstable if at least one of discrete pairs $(k_{y,m},k_{z,n})$ falls
inside the instability region at given frequency, as illustrated in
Fig. \ref{Fig-IntabReg-kykz}a.

At finite radiation corrections and at sufficiently strong in-plane
dissipation the instability region may vanish completely in some frequency
range. The condition for absence of instability can be written as
\begin{align*}
&  \max_{k_{z}}\left[  \frac{k_{z}^{2}}{1\!+\!\nu_{ab}^{2}\omega_{m}^{2}
}\!-\!\frac{\tilde{\delta}_{\omega}}{\omega_{m}}\!+\!\frac{\nu^{2}}
{4\tilde{\delta}_{\omega}\omega_{m}}\right. \\
&  \left.  +\sqrt{\left(
\frac{\tilde{\delta}_{\omega}}{\omega_{m}}+\frac
{\nu^{2}}{4\tilde{\delta}_{\omega}\omega_{m}}\right)
^{2}+\frac{\nu_{k}^{2} }{\omega_{m}^{2}}}\right]  <0
\end{align*}
Finding the maximum leads to the following range of frequency at which there
is no instability at all
\begin{align}
&  \left\vert \frac{|\delta_{\omega}|}{\omega_m}-\frac{W_{ab}}{2}\left(  \frac{\nu_{c}}{\omega_{m}
}+\omega_{m}\nu_{ab}\alpha_{r}\right)  \right\vert \nonumber\\
&  <\!\frac{W_{ab}}{2}\sqrt{\left(  \frac{\nu_{c}}{\omega_{m}}\!+\!\omega
_{m}\nu_{ab}\alpha_{r}\right)  ^{2}\!-\!\frac{\nu^{2}}{\omega_{m}^{2}}}
\label{FreqStabRegion}
\end{align}
with $W_{ab}=\sqrt{1+\omega_{m}^{2}\nu_{ab}^{2}}+\omega_{m}\nu_{ab}$. This
region only exists if inequality $\omega_{m}^{2}\nu_{ab}\alpha_{r}>\nu_{r}$ is
satisfied, leading to the following condition for the in-plane dissipation
$\omega_{m}\nu_{ab}>2\pi/\ln[C/k_{\omega}L_{z}]$.

The largest value of $k_{z}$ at which the instability boundary
crosses $k_{y}=0$, $k_{z0}$, can be found as
\begin{widetext}
\begin{equation}
k_{z0}^{2}
\!=\alpha_{r}\!-\frac{\tilde{\delta}_{\omega}}{\omega_{m}}
-\!\frac{\nu^{2}}{4\tilde{\delta}_{\omega}\omega_{m}}\!+\!\nu_{c}\nu_{ab}
+\!\sqrt{\left[  1+\left(  \omega_{m}\nu_{ab}\right)  ^{2}\right]
\left(
\frac{\tilde{\delta}_{\omega}}{\omega_{m}}+\!\frac{\nu^{2}}{4\tilde{\delta
}_{\omega}\omega_{m}}\right)  ^{2}\!-\!\left[
\frac{\nu_{c}}{\omega_{m} }\!+\omega_{m}\nu_{ab}\left(
\alpha_{r}\!-\!\frac{\tilde{\delta}_{\omega}
}{\omega_{m}}\!+\!\frac{\nu^{2}}{4\tilde{\delta}_{\omega}\omega_{m}}\right)
\right]  ^{2}} \label{kz0}
\end{equation}
\end{widetext}
A finite-height stack is stable with respect to the long-range
deformations if the minimum wave vector
$k_{z,\min}=\pi\ell/N=\pi\lambda_{ab}/L_{z}$ exceeds the maximum
between the two frequency-dependent wave vectors $k_{z,i}$ and
$k_{z0}$ defined by Eqs.\ (\ref{NoInstab-q}) and (\ref{kz0}) leading
to the following criterion
\begin{equation}
    N<\frac{\pi\ell}{\max(k_{z,i},k_{z0})}.
    \label{StbCritN}
\end{equation}
This condition gives to the frequency-height stability diagrams
illustrated in Fig. \ref{Fig-SizeFreqDiag-vab}. As the radiation
corrections $\alpha_{r}$ and $\nu_{r}$ are proportional to the stack
height, these boundaries have to be computed self-consistently. A
simple estimate for the wave vector $k_{z0}$ can be obtained for
weak in-plane dissipation $\omega_{m}\nu_{ab} \ll1$ away from the
resonance at $\tilde{\delta}_{\omega}<0$,
$|\tilde{\delta}_{\omega}|\gg \nu, \omega_{m}\alpha_{r}$. In this
case $k_{z0}^{2}\approx2|\tilde{\delta }_{\omega}|$ meaning that the
finite-height becomes unstable at $|\tilde
{\delta}_{\omega}|/\omega_{m}>(\pi\ell/N)^{2}/2$.

In presence of radiation corrections, two regimes exists depending
on strength of the in-plane dissipation. At small
$\omega_{m}\nu_{ab}$ there is always the critical stack height above
which the system becomes unstable. Coupling to the radiation
decreases this critical height, i.e., it enhances the instability.
This is in contrast to the small-size stacks away from
resonances\cite{BulKoshPRL07}, where the coupling with outside radiation
stabilizes the synchronized state. This behavior changes at large
inplane dissipation. In this case a range of frequencies exists
within which the system remains stable for all stack heights. Also,
in this regime coupling to radiation somewhat increases the critical
stack height away from the resonance.

\subsection{Test of the long-range instabilities with numerical simulations}

\begin{figure}[ptb]
\begin{center}
\includegraphics[width=3.0in]{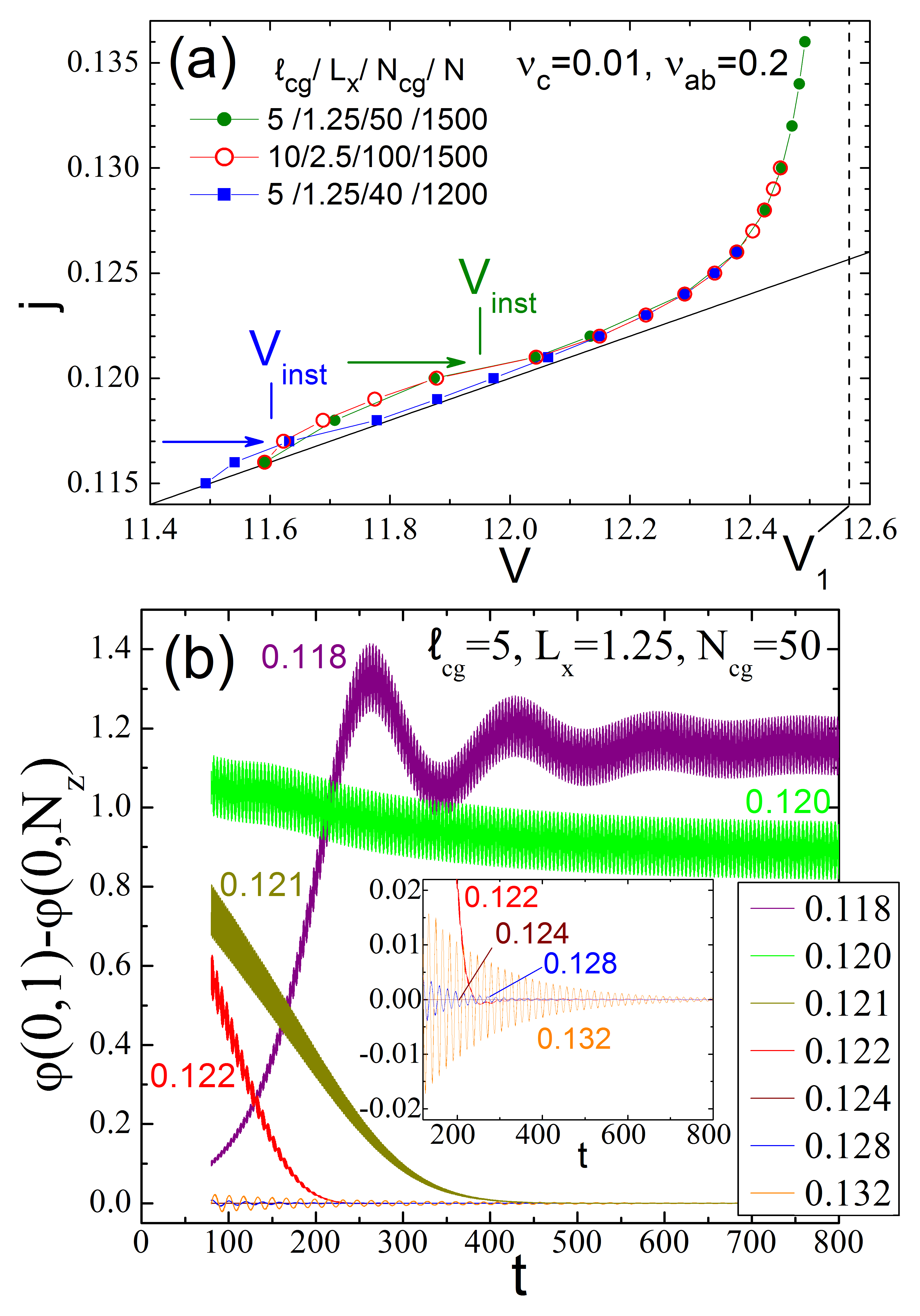}
\end{center}
\caption{Probing the long-range instability using numerical
simulations of the coarse-grained model. (a) The current-voltage
dependences for three sets of parameters below the resonance voltage
$V_1=4\pi$. The horizontal arrows mark onsets of instabilities in
simulations and the vertical bars mark the theoretical estimates
described in the text. The instability leads to appearance of a
small additional bump in the current-voltage dependence. (b) The
time evolution of the difference between the bottom and top phases
at the left edge. One can see that for $j\geq 0.121$ this difference
decays with time indicating stability of the homogeneous state,
while for $j\leq 0.120$ this difference remains finite. The inset
shows the blowup plot of the long-time decay of the phase
differences for stable states. The stripelike appearance of the
curves is due to the rapid oscillations with the Josephson
frequency.} \label{Fig-LongRangeIstSim}
\end{figure}
\begin{figure*}[ptb]
\begin{center}
\includegraphics[width=6.8in]{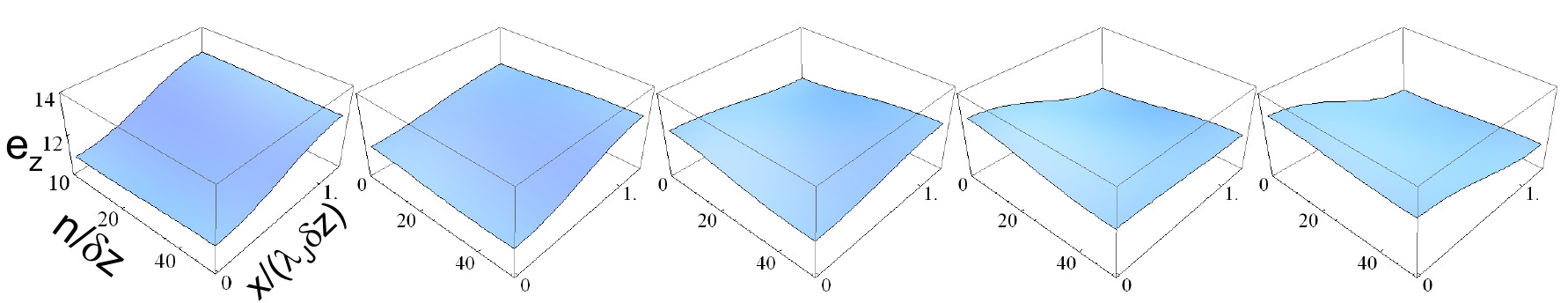}
\end{center}
\caption{Snapshots of the distribution of the electric field for the
steady state in the unstable region for the first set of parameters
in the previous figure at $j=0.12$ over the half period of the
Josephson oscillations. One can see the contribution from the
nonuniform mode.} \label{Fig-SnapshotsElField}
\end{figure*}

We made several approximations in our analytical derivations, which
allowed us to obtain relatively simple criteria for the long-range
stability. To verify validity of these approximations, we checked
some of the analytical results numerically. As suggested by the
representative phase diagrams shown in Fig.\
\ref{Fig-SizeFreqDiag-vab}, the long-range instabilities are only
expected in rather tall stacks with heights $N$ exceeding $1000$
junctions. It is very difficult to simulate such tall stacks
directly. Fortunately, if we only interested in the long-range
instabilities, this is not necessary. For this purpose, we simulated
a coarse-grained model with the step in c-direction $\delta z$
containing many junctions. As demonstrated in the Appendix
\ref{App-Num}, by change of variables, this coarse-grained model can
be reduced to the original model with reduced parameter $\ell$,
$\ell\rightarrow \ell_{\mathrm{cg}}=\ell /\delta z$. This trick
allows us to use the same code with different parameters to probe
the long-range stability of very tall stacks. In this case, the
number of layers in the model $N$ is replaced by the number of
numerical slices $N_{\mathrm{cg}}=N/\delta
z=N\ell_{\mathrm{cg}}/\ell$. In numerics we use the two-dimensional
model which only allows us to check our analytical results in the
simplest situations when the instability is homogeneous in
y-direction ($k_y=0$). We also neglected the layer-charging effect,
$\alpha=0$, and did not take into account the radiation corrections,
$\alpha_r=\nu_r=0$. Having in mind to probe the long-range stability
of the kink state, we use the modulation function
$g(u)=\mathrm{sign}(u-L_x/2)$. With such modulation function the
system is stable with respect to the short-scale perturbations

To probe the long-range stability, we numerically solved the
dynamics equations for increasing transport current in the voltage
range corresponding to the Josephson frequencies close and below the
fundamental resonance. We added small deformation $\delta \varphi
(u,n) \propto \cos (\pi (n-1/2)/N)$ at the beginning of every run
for new value of the current and monitored the time evolution of the
difference $\varphi (0,1)-\varphi (0,N_{\mathrm{cg}})$. Figure
\ref{Fig-LongRangeIstSim} shows results of simulations of the
coarse-grain model, which reveal the long-range instability. Figure
\ref{Fig-LongRangeIstSim}a shows the current-voltage dependences
near the resonance for three sets of parameters: (i)
$\ell_{\mathrm{cg}}\!=\!5$, $L_{x}\!=\!1.25 \lambda_J\delta z$, and
$N_{\mathrm{cg}}\!=\!50$, (ii) $\ell_{\mathrm{cg}}\!=\!10$,
$L_{x}\!=\!2.5 \lambda_J\delta z$, and $N_{\mathrm{cg}}\!=\!50$, and
(iii) $\ell_{\mathrm{cg}}\!=\!5$, $L_{x}\!=\!1.25 \lambda_J\delta
z$, and $N_{\mathrm{cg}}\!=\!40$. In all cases we used the
dissipation parameters $\nu_c=0.01$ and $\nu_{ab}=0.2$. The values
of the total stack heights
$N=N_{\mathrm{cg}}\ell/\ell_{\mathrm{cg}}$ listed in the legend were
obtained assuming $\ell=150$. For all three cases we observe the
long-range instability which leads to appearance of a small
additional bump in the current-voltage dependence. Development of
the instability is illustrated in Fig.\ \ref{Fig-LongRangeIstSim}b
in which we show time evolution of $\varphi (0,1)-\varphi
(0,N_{\mathrm{cg}})$ for different currents for the first set of
parameters. We can see that for  for $j\geq 0.121$ the initial
perturbation decays with time indicating stability of the
homogeneous state, while for $j\leq 0.120$ the perturbation does not
decay. The instability onsets are marked by the horizontal arrows in
Fig.\ \ref{Fig-LongRangeIstSim}a. The sets of parameters (i) and
(ii) correspond to simulations of the same physical system with two
different coarse-graining parameters, $\delta z=30$ and $15$
(assuming $\ell=150$). We observed that in both cases the
instability develops at the same voltage indicating that the
coarse-graining does not influence much the long-range stability.
The set (iii) has the same parameters as the set (i), except for the
smaller height. We see that the instability moved to the lower
voltage, as expected.

We now compare the location of the instability onset with the
analytical predictions. At $k_{y}=0$ the condition of stability is
given by $\pi \ell /N > k_{z0}$ and,  without the radiation
corrections, the formula (\ref{kz0}) for $k_{z0}$ significantly
simplifies,
\begin{equation}
k_{z0}^{2}=2\left( \frac{|\delta _{\omega }|}{\omega
_{1}}-\!\frac{\nu _{c}^{2}}{4|\delta _{\omega }|\omega _{1}}-\nu
_{c}\nu _{ab}\right), \text{ for } \alpha_r,\nu_r=0.
\end{equation}
Moreover, for the parameters we used in simulations the terms with
$\nu_c$ are negligible and with high accuracy we can estimate the
shift from the resonance where the instability is expected as
\begin{equation*}
|\delta _{\omega }|\approx(\omega _{1}/2)(\pi\ell/N)^2=(\omega
_{1}/2)(\pi\ell_{\mathrm{cg}}/N_{\mathrm{cg}})^2.
\end{equation*}
This gives $|\delta _{\omega }|\approx 0.62$,
 $V_{\mathrm{inst}}\!=\!\omega_1-|\delta _{\omega }|\!\approx
\!11.95$ for the first and second sets of parameters and $|\delta
_{\omega }|\!\approx \!0.97$, $V_{\mathrm{inst}}\!\approx \!11.6$
for the third set. These values are shown by the vertical bars in
Fig. \ref{Fig-LongRangeIstSim}a.  We see that in the simulations the
instabilities appear \emph{exactly} where they are predicted
analytically.  This gives us a confidence that the used approximations are legitimate.

With simulations we can go beyond finding the location of the
instability onset. We can also find the finite dynamic state after
the instability develops. To understand the structure of this finite
state,  we present in Fig.\ \ref{Fig-SnapshotsElField} snapshots of
the distribution of the electric field for the first set of
parameters at $j=0.12$. Analyzing these snapshots, we conclude that
the instability leads to the state in which the oscillating phase is
a superposition of two modes,
\[
\theta_{\omega,n}(x)=[\Psi_{1,0}+i\Psi_{1,1}\cos(\pi(n\!-\!1/2)/N)]\cos(\pi
x/L_{x}),
\]
where the amplitude of the nonuniform mode,
$\Psi_{1,1}$, continuously grows starting from zero at the
instability onset.

\section{Summary}

In conclusion, we found that dynamical states synchronized by the
internal cavity resonance are prone to two very different
instabilities. \emph{The short wave-length} instability develops for
states which have regions of negative time-averaged Josephson
coupling. In particular, the homogeneous state in stacks with
modulated Josephson coupling typically has this type of instability.
The homogeneous state in the external magnetic field
$H<\Phi_0/(sL_x)$ has this type of instability close to the
resonance and the instability range widens with decreasing field.
\emph{The long wave-length} instability appears due to the
parametric resonance excitation of the fast modes at finite wave
vectors. The instability criterion depends on the relation between
the damping of the homogeneous mode and modes at finite wave
vectors. Finite-height stacks are stable sufficiently close to the
resonance. The instability region typically shrinks with increasing
inplane dissipation.

\begin{acknowledgments}
I would like to acknowledge many useful discussions with U.\ Welp, L.\
Bulaevskii, X.\ Hu, S.\ Z.\ Lin, K.\ Gray, L.\ Ozyuzer,
K.\ Kadowaki, H.\ Wang, and R.\ Kleiner. This work was supported by
UChicago Argonne, LLC, operator of Argonne National Laboratory, a
U.S. Department of Energy Office of Science laboratory, operated
under contract No. DE-AC02-06CH11357.
\end{acknowledgments}

\appendix
\section{Numerical simulations and coarse-graining procedure \label{App-Num}}
\begin{figure}[ptb]
\begin{center}
\includegraphics[width=2.5in]{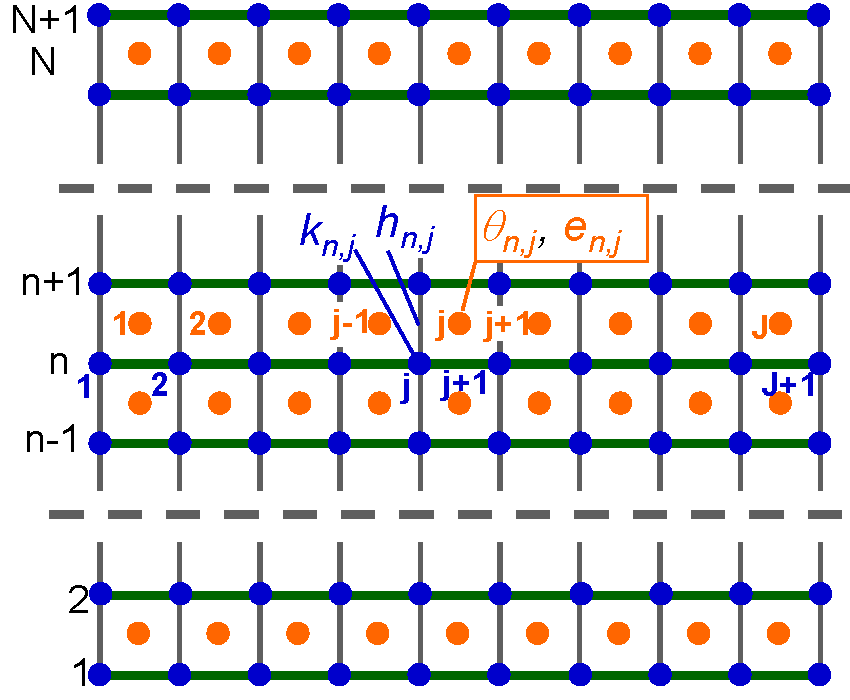}
\end{center}
\caption{Illustration of the staggered grid used for numerical
solution of the dynamic equations (\ref{NumEq1})-(\ref{NumEq4}).}
\label{Fig-Discr}
\end{figure}

For numerical simulations it is convenient to present the dynamic
equations in the form of the time-evolution equations for the
reduced c-axis electric fields ($e_n$), phases ($\varphi_n$),
in-plane supermomenta ($k_n$), and magnetic fields ($h_n$),
\begin{align}
\frac{\partial e_{n}}{\partial\tau} &
=-\nu_{c}e_{n}-g(u)\sin\varphi_{n}
+\frac{\partial h_{n}}{\partial u},\label{NumEq1}\\
\frac{\partial\varphi_{n}}{\partial\tau} &  =e_{n},\\
\nu_{ab}\frac{\partial k_{n}}{\partial\tau} &  =-\left[  k_{n}+h_{n}
-h_{n-1}\right],\\
h_{n} &  =\ell^{2}\left(  \frac{\partial\varphi_{n}}{\partial
u}-k_{n+1} +k_{n}\right).\label{NumEq4}
\end{align}
The units in these equations are different from units used for
analytical calculations: unit of length is the Josephson length
$\lambda_J$, unit of supermomentum is $1/\lambda_J$, unit of
magnetic field is $\Phi_0/(2\pi\gamma\lambda^2)$, and unit of
electric field is $\Phi_0\omega_p/(2\pi cs)$. All parameters are
assumed to be y-independent. Therefore we only probe instabilities
uniform in this direction. We also neglected the layer-charging
effect, $\alpha=0$. Above equations are solved for stack containing
$N$ junctions with $0<u<L_x$ assuming simple non-radiative boundary
conditions at the edges, $k_n=0$, $\partial \varphi_n/\partial u=\mp
I/2\ell^2$ at $u=0,L_x$, where $I=j L_x$ is the total transport
current

Solution of these equations is implemented using the following
implicit numerical scheme:
\begin{widetext}
\begin{itemize}
\item Space and time discretizations are performed using a staggered grid. For
coordinate, $\varphi_{n}(u,\tau)$ and $e_{n}(u,\tau)$ are defined at
the points $u=(j-1/2)d_{u}$, while $k_{n}(u,\tau)$ and
$h_{n}(u,\tau)$ are defined at the points $u=jd_{u}$, see Fig.\
\ref{Fig-Discr}. For time, $e_{n}$ is defined at $\tau=md_{\tau}$
while $\varphi_{n}$, $k_{n}$, and $h_{n}$ are defined at
$\tau=(m+1/2)d_{\tau}$.
\begin{align*}
\varphi_{n,j}^{m+1/2} &  =\varphi_{n}\left[
(j-1/2)d_{u},(m+1/2)d_{\tau}\right] ,\ e_{n,j}^{m}
=e_{n}\left[  (j-1/2)d_{u},md_{\tau}\right]  \text{ }1<j<J\\
k_{n,j}^{m+1/2} &  =k_{n}\left[  (j-1)d_{u},(m+1/2)d_{\tau}\right]
,\ h_{n,j}^{m+1/2}=h_{n}\left[ (j-1)d_{u},(m+1/2)d_{\tau}\right]  ,\
1<j<J+1
\end{align*}

\item We discretize equations as
\begin{align}
\frac{e_{n,j}^{m+1}-e_{n,j}^{m}}{d_{\tau}} &  =-\nu_{c}\frac{e_{n,j}
^{m+1}+e_{n,j}^{m}}{2}-g_{j}\sin\varphi_{n,j}^{m+1/2}+\frac{h_{n,j+1}
^{m+1/2}-h_{n,j}^{m+1/2}}{d_{u}}\label{Eqe}\\
\frac{\varphi_{n,j}^{m+3/2}-\varphi_{n,j}^{m+1/2}}{d_{\tau}} &
=e_{n}
^{m+1}\label{EqDiscPhase}\\
\frac{k_{n,j}^{m+3/2}-k_{n,j}^{m+1/2}}{d_{\tau}} &
=-\frac{1}{\nu_{ab} }\left[
\frac{k_{n,j}^{m+3/2}+k_{n,j}^{m+1/2}}{2}+\frac{h_{n,j}
^{m+3/2}+h_{n,j}^{m+1/2}}{2}-\frac{h_{n-1,j}^{m+3/2}+h_{n-1,j}^{m+1/2}}
{2}\right]  \label{Eqk}\\
h_{n,j}^{m+3/2} &  =\ell^{2}\left(
\frac{\varphi_{n,j}^{m+3/2}-\varphi
_{n,j-1}^{m+3/2}}{d_{u}}-k_{n+1,j}^{m+3/2}+k_{n,j}^{m+3/2}\right).
\label{Eqh}
\end{align}

\item The first two equations allow for direct time advance of $e_{n,j}$ and
$\varphi_{n,j}$
\begin{align*}
e_{n,j}^{m+1} &  =\left( \frac{1}{d_{\tau}}+\frac{\nu_{c}}{2}\right)
^{-1}\left[  \left( \frac{1}{d_{\tau}}-\frac{\nu_{c}}{2}\right)
e_{n,j}
^{m}-g_{j}\sin\varphi_{n,j}^{m+1/2}+\frac{h_{n,j+1}^{m+1/2}-h_{n,j}^{m+1/2}
}{d_{u}}\right]  \\
\varphi_{n,j}^{m+3/2} &  =\varphi_{n,j}^{m+1/2}+d_{\tau}e_{n}^{m+1}
\end{align*}

\item Substitution of $h_{n,j}^{m+3/2}$ and $h_{n-1,j}^{m+3/2}$ from
Eq.\ (\ref{Eqh}) into Eq.\ (\ref{Eqk}) leads to the tridiagonal
linear system for $k_{n,j}^{m+3/2}$,
\begin{align*}
&  \ell^{2}k_{n+1,j}^{m+3/2}-\left(
1+\frac{2\nu_{ab}}{d_{\tau}}+2\ell^{2}\right)
k_{n,j}^{m+3/2}+\ell^{2}k_{n-1,j}^{m+3/2}\\
&  =\left(  1-\frac{2\nu_{ab}}{d_{\tau}}\right)
k_{n,j}^{m+1/2}+\ell^{2}
\frac{\varphi_{n,j}^{m+3/2}-\varphi_{n,j-1}^{m+3/2}-\varphi_{n-1,j}
^{m+3/2}+\varphi_{n-1,j-1}^{m+3/2}}{d_{u}}+h_{n,j}^{m+1/2}-h_{n-1,j}
^{m+1/2}
\end{align*}
for $n=2,\ldots,N$ with $k_{1,j}^{m+3/2}=0;\ k_{N+1,j}^{m+3/2}=0$.
Solving this system, we advance $k_{n,j}$

\item After finding $k_{n,j}^{m+3/2}$, we update $h_{n,j}^{m+3/2}$ using Eq.\ (\ref{Eqh}).
\end{itemize}
\end{widetext}

The long-range instabilities are only expected for very tall stacks
$N>1000$ which are very difficult to simulate directly. To probe
these instabilities, we use the coarse-grained model. Assuming that
the perturbations are smooth in z-direction we introduce a
discretization step $\delta z$ containing many junctions, $\delta
z\gg1$, and write coarse-grained equations only for junctions with
$n=\delta z m$,
\begin{align*}
\frac{\partial e_{m}}{\partial\tau} &  =-\nu_{c}e_{m}-g(u)\sin\varphi_{m}%
+\frac{\partial h_{m}}{\partial u},\\
\frac{\partial\varphi_{m}}{\partial\tau} &  =e_{m},\\
\frac{\partial k_{m}}{\partial\tau} &  =-\frac{1}{\nu_{ab}}\left[  k_{m}%
+\frac{h_{m}-h_{m-1}}{\delta z}\right]  ,\\
h_{m} &  =\ell^{2}\left(  \frac{\partial\varphi_{m}}{\partial
u}-\frac {k_{m+1}-k_{m}}{\delta z}\right).
\end{align*}
Transforming variables, $u=\delta z\tilde{u}$,
$h_{m}=\tilde{h}_{m}\delta z$, we arrive to original equations
(\ref{NumEq1})-(\ref{NumEq4}) with replacements
$u\rightarrow\tilde{u}$, $h_{m}\rightarrow\tilde{h}_{m}$, and
$\ell\rightarrow\ell_{\mathrm{cg}}=\ell/\delta z=\lambda/\delta z
s$. Therefore, \emph{simulations of the same model with the smaller
parameter} $\ell$ \emph{is equivalent to coarse-graining} and allows
us to explore the long-range instabilities in very tall stacks.
Renormalization implies that the stack width is now measured in
units $\lambda_J \delta z$. The effective stack height is given by
$N=\delta z N_{\mathrm{cg}}$, where $N_{\mathrm{cg}}$ is the total
number of the c-axis slices in the coarse-grained model. Having in
mind to probe the long-range stability of the kink state, we use the
modulation function $g(u)=\mathrm{sgn}(u-L_x/2)$. With such
modulation function the system is stable with respect to the
short-scale perturbations.


\begin{thebibliography}{99}
\bibitem{footnoteSelfField}We neglect small correction to the phase coming
from self-field of transport current. It has a negligible influence
on our results.
\bibitem{footnote}The mesas, in which resonance terahertz emission was
detected in Ref. \onlinecite{LutfiSci07}, were fabricated on the top
of bulk crystals. In this case, the dominating dissipation channel
is leaking of radiation into the base crystal \cite{KoshBulLT25}.
Stability analysis in this case is more complicated and we are not
doing it in this paper.
\end{thebibliography}
\end{document}